\documentclass[aps,onecolumn,notitlepage,superscriptaddress,floatfix,showkeys]{revtex4-1}
\usepackage{amsmath,amssymb,amsfonts}   
\usepackage{lineno,hyperref}
\usepackage{graphicx}
\usepackage{epstopdf}
\modulolinenumbers[1]

\renewcommand{\thesection}{\arabic{section}}
\renewcommand{\thesubsection}{\thesection.\arabic{subsection}}
\renewcommand{\thesubsubsection}{\thesubsection.\arabic{subsubsection}}

\makeatletter
\renewcommand{\p@subsection}{}
\renewcommand{\p@subsubsection}{}
\makeatother

\begin{document}


\title{Consequences of glacial cycles for magmatism and carbon transport at mid-ocean ridges}

\author{Nestor G. \surname{Cerpa}} 
\affiliation{Department of Earth Sciences, University of Oxford, South Parks Road, Oxford, OX1 3AN, UK.}
\affiliation{University of Montpellier, G{\'e}osciences Montpellier, Place Eug\`{e}ne Bataillon, 34090 Montpellier, France.}
\author{David W. \surname{Rees Jones}}
\affiliation{Department of Earth Sciences, University of Oxford, South Parks Road, Oxford, OX1 3AN, UK.}
\affiliation{University of Cambridge, Bullard Laboratories, Department of Earth Sciences, Madingley Road, Cambridge, CB3 0EZ, UK.}
\affiliation{University of St Andrews, School of Mathematics and Statistics, Mathematical Institute, North Haugh, St Andrews, KY16 9SS, United Kingdom}
\author{Richard F. \surname{Katz}}
\affiliation{Department of Earth Sciences, University of Oxford, South Parks Road, Oxford, OX1 3AN, UK.}

\begin{abstract}
    Magmatism and volcanism transfer carbon from the solid Earth into the climate system. This transfer may be modulated by the glacial/interglacial cycling of water between oceans and continental ice sheets, which alters the surface loading of the solid Earth.  The consequent volcanic-carbon fluctuations have been proposed as a pacing mechanism for Pleistocene glacial cycles. This mechanism is dependant on the amplitude and lag of the mid-ocean ridge response to sea-level changes. Here we develop and analyse a new model for that response, eliminating some questionable assumptions made in previous work. Our model calculates the carbon flux, accounting for the thermodynamic effect of mantle carbon: reduction of the solidus temperature and a deeper onset of melting. We analyse models forced by idealised, periodic sea level and conclude that fluctuations in melting rate are the prime control on magma and carbon flux. We also discuss a model forced by a reconstruction of eustatic sea level over the past 800~kyr. It indicates that peak-to-trough variations of magma and carbon flux are up to about 20\% and 10\% of the mean flux, respectively. Peaks in mid-ocean ridge emissions lag peaks in sea-level forcing by less than about $20$~kyr and the lag could well be shorter. The amplitude and lag are sensitive to the rate of melt segregation.  The lag is much shorter than the time it takes for melt to travel vertically across the melting region.
\end{abstract}


\maketitle

\section{Introduction}

Terrestrial climate has changed dramatically between glacial and interglacial periods during the Pleistocene epoch. Continental ice sheets grow during glacial periods, causing a drop of up to 130~m in eustatic sea level; this decrease is recovered during interglacials \citep{lisiecki2005pliocene, bintanja2005modelled, waelbroeck2002sea, siddall2010changes}. The shift of mass-loading between continents and oceans affects subaerial and submarine volcanism \citep{Jull1996deglaciation} and its consequent carbon transfer from the solid Earth into the atmosphere/ocean \citep{huybers2009feedback}. During the Pleistocene, the climate system has varied on time-scales associated with Milankovitch orbital periods \citep{hays1976variations}, indicating that glacial cycles are externally forced by variations in insolation (and its distribution over latitude). \citet{huybers2009feedback} argued that glacial--volcanic coupling creates an internal amplification of climate variations. They later   hypothesised \citep{huybers2017delayed} that this feedback explains why glacial cycles of the last $\sim$700~kyr have a period of 100~kyr whereas the dominant Milankovitch forcing has a periodicity $\sim$41~kyr \citep{abe2013insolation}. However, the climate-feedback hypothesis depends on the amplitude of the glacial/interglacial fluctuation in carbon emissions and its lag with respect to changes in sea level.  The present manuscript aims to develop a rigorous theory to predict the amplitude and lag.

Eighty percent of global volcanism occurs at mid-ocean ridges (MOR), where tectonic-plate divergence induces upwelling of the underlying mantle. Here, magma is produced by decompression melting, a quasi-isentropic process \citep{mckenzie1988volume, langmuir1992petrological}. Up to about 20\% partial melting occurs in the upper $\sim$100~km of the mantle, within a zone that extends $\sim$100~km on each side of the plate boundary \citep{katz2008magma, keller2017volatiles}. The melt segregates under its buoyancy, which supplies magma to the ridge axis and forms the oceanic crust. Because melting is driven by pressure change, and because variations of sea level affect the static pressure below, it is plausible that glacial cycles modulate magmatic production \citep{huybers2009feedback, lund2011does}. Indeed a simple estimate and more detailed calculations by \citet{crowley2015glacial} indicate that crustal thickness could change by $\sim$10\% \citep[see also][]{tolstoy2015mid}. \citet{crowley2015glacial} argued that such fluctuations provide a mechanism for the formation of abyssal hills with Milankovitch periodicity. This idea is controversial; the standard theory holds that seafloor topography is tectonically controlled \citep[e.g.,][]{olive2015sensitivity}. \citet{lund2011does} hypothesized that sea-level variations impacted the hydrothermal activity and the geochemistry of seawater, which has found some support in sedimentological records \citep{lund2016enhanced,costa2017hydrothermal}. 

The mantle contains about 10$^4$ times more carbon than the atmosphere and ocean combined \citep{Sleep2001cycling, dasgupta2010deep, hirschmann2018comparative}. This carbon is transferred from the mantle into the ocean-atmosphere system by volcanism and returns to the mantle in subduction zones \citep{Zahnle2002cycling, kelemen2015reevaluating}. At mid-ocean ridges, \citet{hirschmann2018comparative} estimated that 120$\pm$26~Mt~CO$_2$/yr is extracted and emitted from the solid Earth; the majority of studies cited by \cite{hirschmann2018comparative} have estimates that differ by within a factor of $\sim$3. The carbon flux could be modulated by variations in MOR magmatism during glacial cycles. \citet{burley2015variations} hypothesised that the key coupling mechanism is the pressure-driven variation in the depth of the onset of silicate melting. A drop in sea level reduces the static pressure and hence deepens the onset of first silicate melting.  The downward motion of this boundary enhances the flux of carbon into the melting region. The opposite occurs when sea level rises. \citet{burley2015variations} predicted that 100 m changes in sea level with periodicity in the range 20--100~kyr would produce fluctuations up to 10\% in the emission rate of CO$_2$. Their hypothesised mechanism creates a lag of 50--80~kyr between a peak in the forcing and a peak CO$_2$ emission rate. This delay arises from the time required for carbon-enriched (or depleted) melts to travel from the base of the silicate melting region to the surface. \citet{huybers2017delayed} found that lags of 10--50~kyr are conducive to a negative feedback that would pace glacial cycles at a frequency of 1/100~kyr$^{-1}$ during the Pleistocene epoch. 

\citet{burley2015variations} invoked significant assumptions and approximations in the development of their model. Two assumptions are particularly relevant here. First, they neglected the fluctuations of melting rate, porosity, and melt-transport speed that arise from sea-level variation \citep{lund2011does, crowley2015glacial}.  Second, they neglected the effect of carbon on mantle melting, which is to drastically lower the solidus temperature at constant pressure or, equivalently, to increase the pressure of first melting at constant entropy \citep[e.g.,][]{gaetani1998influence, dasgupta2006melting}. Indeed, many studies have shown that even low concentrations of volatiles ($\sim$100~ppm) can induce the formation of low-degree melts at depths far greater than that of the anhydrous solidus temperature \citep[e.g.,][]{dasgupta2013carbon,dasgupta2018volatile}. 

The present study aims to develop a more robust mathematical theory by removing simplifying assumptions made by previous work \citep{crowley2015glacial, burley2015variations}. Our models are more complex than previous work in terms of the thermodynamic effect of CO$_2$ and consistently modelling all the consequences of pressure fluctuations. In particular, our theory accounts for both the pressure-induced variations in the onset of (volatile-enriched) melting and also the pressure-induced variations in the melting rate. In this context, we show that the melting-rate variations created by oscillating sea level cause melt-flux variations, and that these are the primary cause of variations in CO$_2$ emissions. Variations in the carbon concentration are secondary and variations in the onset depth of melting are inconsequential. As in previous work, we quantify the model sensitivity to sea-level change in terms of the admittance, which is defined as the amplitude ratio of response to forcing, as a function of frequency.  We obtain a similar admittance of carbon flux as did \citet{burley2015variations}, but we find that carbon emissions lag the causative changes in sea level by less than 20~kyr, assuming that melt ascent at mid-ocean ridges is no slower than 1~m/yr. Faster melt extraction corresponds to shorter lags. The lag is much shorter than the melt travel time because extra melts are generated throughout the column.

The manuscript is structured as follows. In Section~\ref{sec:Methods}, we describe the physical model, the governing equations and the mathematical strategy used to analyse them. In Section~\ref{sec:Results}, we describe the results. We present the steady and time-dependent model predictions of porosity, carbon concentration, melt flux and carbon flux. We determine the admittance and lag of the fluxes as a function of the period of the sea-level cycle. In Section~\ref{sec:Discussion}, we compare our results to previous models \citep{crowley2015glacial, burley2015variations} so as to isolate and discuss the differences. Finally, we consider a calculation of melt and carbon flux variations arising from a reconstruction of sea level over the past 800~kyr.  Appendices provide details of the derivations and analyses that support our findings.

\section{Methods} \label{sec:Methods}
\begin{figure}[!htbp]
    \centering 
    \includegraphics[width =0.4\linewidth]{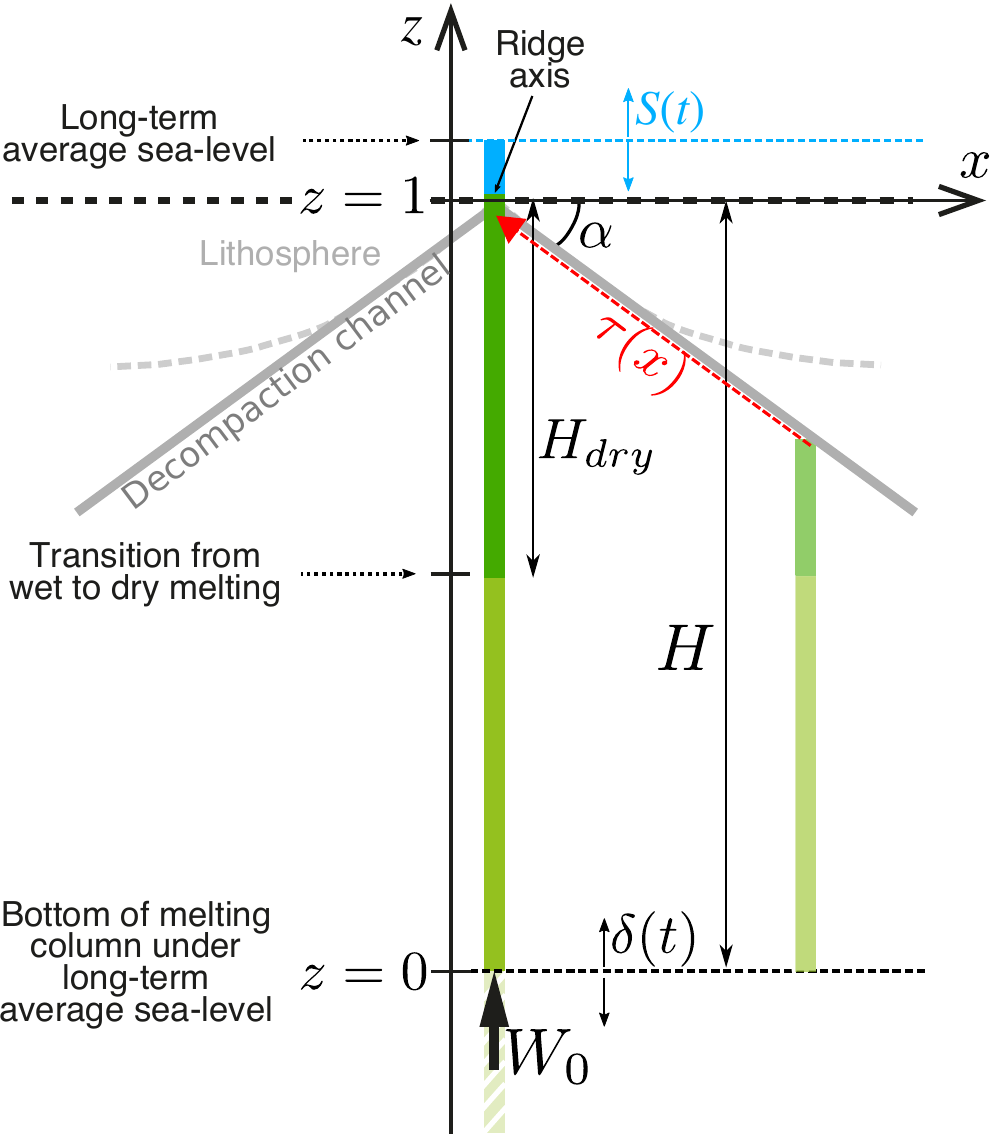}
    \caption{Sketch of the melting region beneath the ridge axis and sea water above it. The green and dark green regions represent the wet- (i.e., carbon-rich) and dry-melting regimes, respectively. The dimensional mantle upwelling rate $W_0$ is represented by a thick black arrow. The dimensional total depth of the melting region $H$ and dry melting region $H_\mathrm{dry}$ are indicated. All other quantities are non-dimensional, as described in the main text. We also sketch the triangular melting region and our pseudo-2d model that is based on combining a series of 1d columns. The base of the triangle corresponds to the depth at which mantle crosses the (wet) solidus temperature. The sides of the triangle correspond to the decompaction channel along which distal melts are focused toward the ridge axis. An off-axis melting column shown at a distance $x$ from the axis empties into the decompaction channel. The transport time of the magma that enters into the channel at $x$ is $\tau(x)$. See section \ref{sec:2d-description} for details.}
    \label{fig:sketch}
\end{figure}

Here, a physical model is developed in mathematical terms to quantify the effect of sea-level variations on the melt and carbon flux to the ocean--atmosphere system. We generalize a standard, steady-state, one-dimensional melting-column model of the upwelling mantle \citep{Ribe1985}. The crucial generalization is to account for the time-dependent melting caused by a time-dependent sea level. We also calculate the evolution of the carbon concentration and account for its thermodynamic effect \citep{dasgupta2006melting}. Figure~\ref{fig:sketch} is a schematic diagram of the melting column, depicting the effect of variable sea level and carbon on the depth at which the mantle crosses the solidus temperature. A pseudo-2d model is constructed from a series of columns. 

\subsection{Physical model}
The physical model consists of a mechanical model of two-phase flow, a thermo-petrological model of melting, and a chemical model of carbon transport. 

\subsubsection{Mechanical model}
Deep beneath a mid-ocean ridge, the mantle upwells at a rate $W_0$. At some depth (pressure) during this rise, its temperature reaches and exceeds the solidus temperature; at this point, it begins partial melting.  The melt produced has a lower density than the residual solid and so rises buoyantly. This difference is small relative to the mean density, and hence we make a Boussinesq approximation and neglect density differences between phases, except when calculating the buoyancy. The volume fraction (or equivalently mass fraction) occupied by the melt is $\phi$. To quantify the key physical controls on the system, we derive a set of non-dimensional governing equations.  In these equations, we non-dimensionalize lengths with the height of the melting column $H$, velocities with $W_0$, and time with $H/W_0$.
However, we keep thermodynamic variables (temperature $T$ and pressure $P$) as dimensional quantities.

Bulk mass conservation (averaged across the two phases) requires that the bulk, one-dimensional flux is constant and equal to the mantle upwelling rate.  In non-dimensional units, bulk mass conservation is
\begin{equation} \label{eq:bulk_mass}
    \phi w +(1-\phi)W = 1,
\end{equation}
where $w$ is the liquid velocity and $W$ is the solid velocity. This expression allows us to determine the liquid flux $Q \equiv \phi w$ in terms of the solid flux. 

The volumetric melting rate, hereafter referred to as the melting rate, can be determined by considering mass conservation in the solid phase and is given by 
\begin{equation} \label{eq:solid_mass}
    \Gamma = \frac{\partial \phi}{\partial t} - \frac{\partial  }{\partial z}\left[(1-\phi)W\right].
\end{equation}

Melt segregates from the solid under gravity $g$ because it is less dense by an amount $\Delta \rho$. The melt has a viscosity $\mu$ and, at the scale of mantle grains, it inhabits an interconnected network of pores with permeability $k \phi^n$ \citep[e.g.,][]{miller2014experimental, rudge18}.  The prefactor $k$ is a constant. The exponent $n$ determines the sensitivity of permeability to porosity and has been estimated to be $2\leq n\leq 3$.  Then the melt segregation velocity is
\begin{linenomath*}
    \begin{align} 
    \phi(w-W) W_0 &= \frac{ \Delta \rho g k \phi^n (1-\phi)}{\mu},  \nonumber \\
    \Rightarrow \quad 1-W &= \mathcal{Q}  \phi^n (1-\phi), \label{eq:darcy2}
    \end{align}
\end{linenomath*}
where the implication follows from equation~\eqref{eq:bulk_mass}. The parameter 
\begin{equation}
    \mathcal{Q} \equiv  \frac{ \Delta \rho g k }{\mu W_0}
\end{equation}
is the ratio of the rate of buoyancy-driven magma segregation to the rate of mantle upwelling.  Thus, given the porosity, the solid flux can be determined from equation~\eqref{eq:darcy2} and then the liquid flux from equation~\eqref{eq:bulk_mass}.

In this formulation, we have neglected the isotropic stress associated with compaction. Compaction stress typically varies on length scales much shorter than $H$ and plays a role only in narrow boundary layers \citep{Ribe1985}. However, compaction stresses can give rise to transient features called compaction waves or magmons \citep{Scott1984, Richter1984}. These could potentially interact with the time-dependence caused by sea-level variation, modifying the rate of chemical transport \citep{Jordan2018}. 

\subsubsection{Thermo-petrological model}
The thermo-petrological model is used in concert with energy conservation to determine the melting rate. A simple parameterisation \citep{reesjones18epsl,bo2018melting} of solidus temperature that increases with pressure and decreases with carbon content is given by the linear relationship
\begin{equation} \label{eq:solidus}
    T = T_0 + \gamma^{-1} (P-P_0) + M c_0 (1-c_s),
\end{equation}
where $T$ is the solidus temperature, $P$ is total pressure, $\gamma$ is the slope of the carbon-free solidus, $M$ is the dependence of the solidus temperature on the carbon concentration $c_0$ of the unmelted mantle, and $c_s$ is the carbon concentration of the solid phase after it has been scaled by $c_0$. The last definition means that the $c_s=1$ before the onset of melting. $T_0$ and $P_0$ are the temperature and pressure at the onset of melting.

We introduce a dimensionless, time-dependent sea level $S(t)$ with time mean $S_0$; we introduce an independent vertical coordinate $z$. The origin $z=0$ is taken to be the bottom of the melting column, the depth at which upwelling mantle achieves the solidus temperature in the absence of sea-level variation ($S=S_0$). All lengths are scaled by $H$. 

The total pressure is affected by sea level.  Neglecting dynamic (compaction) pressure, the total pressure within the melting column is
\begin{equation} \label{eq:P_lith}
    P = P_0 -\rho g H z + \rho_w g H(S-S_0).
\end{equation}
Deviations of sea level from $S_0$ lead to variations in the depth of the onset of mantle melting at
\begin{equation} \label{eq:delta_def}
    z=\delta(t) \equiv \frac{\rho_w}{\rho}(S - S_0).
\end{equation}

Energy conservation is given by a temperature equation that accounts for advective transport and latent heat release \citep{Ribe1985}
\begin{equation} \label{eq:heat}
  \frac{\mathrm{D} T}{\mathrm{D} t} = -\frac{L}{c_p} \Gamma,
\end{equation}
where $L$ is the latent heat and $c_p$ is the specific heat capacity. Note that $\mathrm{D}/\mathrm{D}t \equiv \partial /\partial t + \partial /\partial z$ is a Lagrangian derivative accounting for bulk advection, i.e., advection by both the solid and liquid phases. We do not consider the effect of compressibility, which is relatively small. We substitute equations~\eqref{eq:solidus} and \eqref{eq:P_lith} into \eqref{eq:heat} to obtain the melting rate
\begin{equation} \label{eq:melting_rate}
    \Gamma = {\underbrace{\vphantom{\frac{d \delta}{d t}}\Gamma^*}_{\Gamma_0}} \,{\underbrace{- \Gamma^* \frac{d \delta}{d t}}_{\Gamma_P}} \,{\underbrace{+ \mathcal{M} \frac{\mathrm{D} c_s}{\mathrm{D} t}}_{\Gamma_c}},
\end{equation}
where the dimensionless parameters are defined by
\refstepcounter{equation}
\[
    \Gamma^* \equiv \frac{\rho g H c_p}{\gamma L}, \qquad  
    \mathcal{M} \equiv \frac{c_p M c_0}{L} .
    \eqno{(\theequation{\mathit{a},\mathit{b}})} 
\]
$\Gamma^*$ is the isentropic, dry melt productivity and $\mathcal{M}$ is the significance of carbon for melting. The latter is the ratio of the sensible heat associated with the solidus-depression caused by carbon to the latent heat. Dry melting corresponds to $\mathcal{M} = 0$; carbonated melting to $\mathcal{M} > 0$. 
Equation~\eqref{eq:melting_rate} shows that there are contributions to the melting rate from standard decompression melting $\Gamma_0$, from the pressure change associated with sea-level variation $\Gamma_P$, and from the effect of carbon on melting $\Gamma_c$. Sea-level variation has a direct effect on melting since
\begin{equation}
    \Gamma_P \equiv - \Gamma^* \frac{d \delta}{d t} = - \Gamma^*  \frac{d}{dt} \left[\frac{\rho_w}{\rho}\left(S-S_0\right)\right].
\end{equation}
Sea-level variation also has an indirect effect on the melting rate by changing the concentration of carbon, a mechanism captured in the term $\Gamma_c$. 

The behaviour of the melting rate has a crucial transition between low-degree, carbonated melting and dry melting. This occurs when $z=\mathcal{M}/\Gamma^*$. Indeed, we can make an alternative interpretation of $\mathcal{M}$ in terms of the total depth $H$ of the melting region (the carbonated solidus depth)  relative to the dry solidus depth $H_\mathrm{dry}$. In particular
\begin{equation} \label{eq:M(Gamma)}
    \mathcal{M} = \Gamma^*\left(1-\frac{H_\mathrm{dry}}{H} \right).
\end{equation}
Note that \mbox{$H_\mathrm{dry} = {\gamma \Delta T}/{\rho g}$}, where $\Delta T$ is the difference between the mantle potential temperature and the solidus temperature at the top of the melting column \citep{Ribe1985}. The maximum degree of melting $F_\mathrm{max}$ is equal to the liquid flux at the top of the column \citep{Ribe1985}. This satisfies, to an excellent approximation at small porosity, 
\begin{equation}
    F_\mathrm{max} = \Gamma^* - \mathcal{M}.
\end{equation}
Thus given estimates for $H_\mathrm{dry}$, $H$ and $F_\mathrm{max}$, the dimensionless parameters $\Gamma^*$ and $\mathcal{M}$ can be uniquely determined (see Appendix~\ref{app:steady} for details). 

\subsubsection{Chemical model}
The final part of the model captures the transport of carbon. Diffusion and dispersion of carbon are slow compared to advection, so conservation of carbon in the two phases is given by
\begin{equation} \label{eq:chemistry1}
     \frac{\partial }{\partial t} \left[\phi c_l +(1-\phi) c_s \right] + \frac{\partial }{\partial z} \left[\phi w c_l +(1-\phi) W c_s \right] = 0,
\end{equation}
where $c_l$ is the dimensionless concentration of carbon in the melt (again scaled by $c_0$). 

We assume that carbon behaves as an incompatible element \citep[e.g.][]{rosenthal2015experimental} that partitions preferentially into the melt according to
\begin{equation} \label{eq:partition}
    c_s = D_c \, c_l, 
\end{equation}
where $D_c$ is a partition coefficient for carbon. Equation~\eqref{eq:partition} allows us to eliminate $c_l$ in equation~\eqref{eq:chemistry1} and obtain a single equation for carbon transport.

\subsubsection{Boundary conditions}
At the depth of the onset of melting $z=\delta(t)$, there is no porosity and the carbon concentration is that of the far-field, upwelling mantle. Thus appropriate boundary conditions are
\refstepcounter{equation}
\[
    \phi = 0, \quad  
    c_s = 1 \qquad (\text{at } z=\delta) .
    \eqno{(\theequation{\mathit{a},\mathit{b}})} \label{eq:bcs1}
\]

\subsubsection{Governing equations}
To synthesise the model components above, we combine the mechanical, thermal and chemical equations to obtain a coupled system. We eliminate the liquid and solid fluxes using equations~\eqref{eq:bulk_mass} and \eqref{eq:darcy2} and the melt rate using equations~\eqref{eq:solid_mass} and \eqref{eq:melting_rate}. Then the system of equations for the evolution of porosity and carbon concentration is
\begin{linenomath*} \begin{subequations} \label{eq:full}
\begin{align}
    &\frac{\partial \phi}{\partial t} 
    +   \frac{\partial Q}{\partial z} =  \Gamma^* - \Gamma^* \frac{d \delta}{d t} +  \frac{\mathrm{D} c}{\mathrm{D} t}, \label{eq:full-a} \\
    & D \frac{\mathrm{D} c}{\mathrm{D} t}  
    + \frac{\partial}{\partial t} \left(c \phi \right)
    + \frac{\partial}{\partial z} \left(c Q \right) 
    = 0, \label{eq:full-b}
\end{align}
\end{subequations} \end{linenomath*}
where we simplified the notation by defining a scaled carbon concentration and effective partition coefficient, respectively,
\refstepcounter{equation}
\[
    c \equiv \mathcal{M}c_s, \qquad  
    D \equiv D_c/(1-D_c).
    \eqno{(\theequation{\mathit{a},\mathit{b}})} 
\]
The liquid flux $Q$ depends on porosity according to
\begin{equation}
    Q(\phi) =\mathcal{Q} \phi^n (1-\phi)^2 + \phi.
\end{equation}
We also evaluate the carbon flux in the melt, $Q_c$, which is defined by
\begin{equation}
    Q_c \equiv \phi w c_l = \frac{Q c}{\mathcal{M} D_c}.
\end{equation}

\subsection{Decomposition into steady and fluctuating components}
\begin{figure}[t]
    \centering 
    \includegraphics[width =0.95\linewidth]{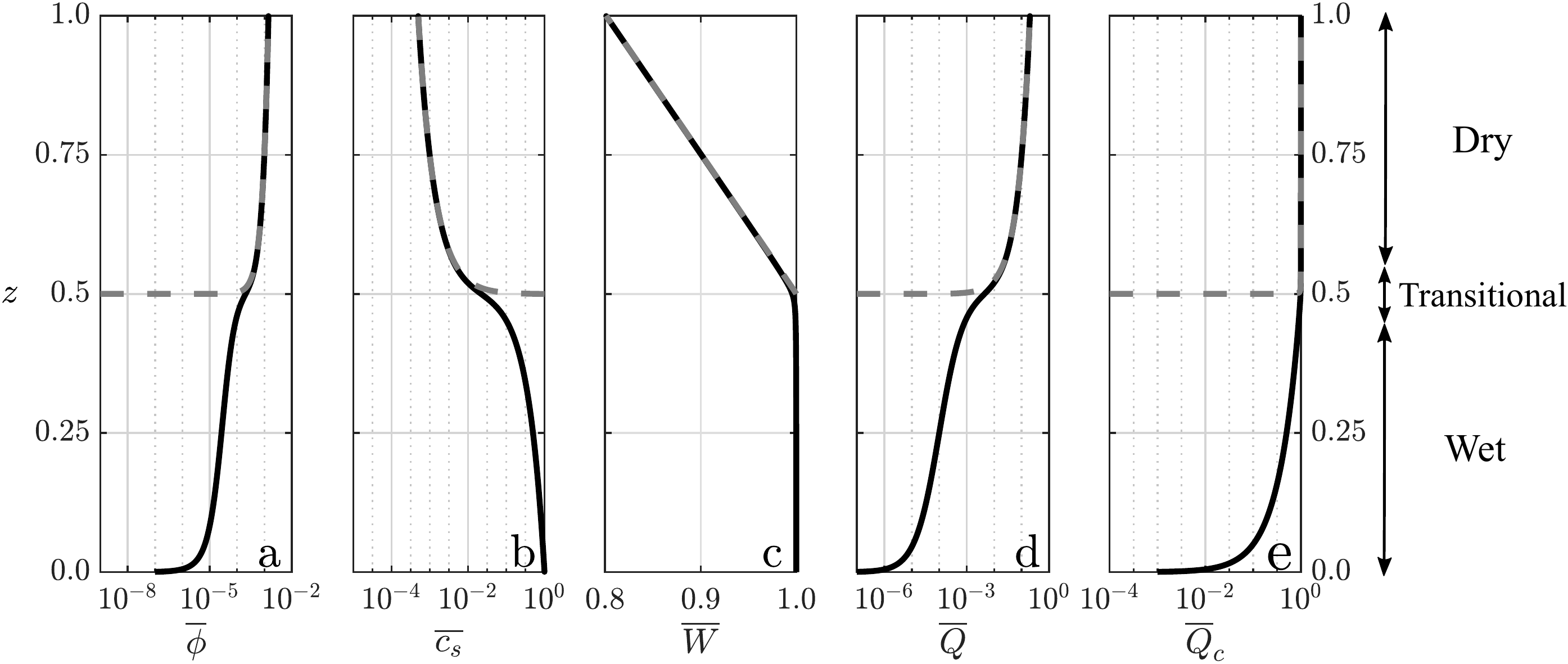}
    \caption{Steady column variables: porosity (a), carbon concentration in solid (b), solid velocity (c), melt flux (d), and carbon flux (e). The solid curves show a wet-melting model; the dashed curves show an otherwise equivalent dry model. As described in the main text, the melting column is divided into three regions: `wet', `transitional' and `dry'. For the parameters used (table~\ref{tab:parameters}), the dry solidus is at $z=0.5$.} 
    \label{fig:base}
\end{figure}

The melting column has a steady-state, mean behaviour in the absence of sea-level fluctuations. Figure~\ref{fig:base} shows a typical steady-state solution of the governing equations. Table~\ref{tab:parameters} gives the standard set of parameters we use in calculations unless otherwise stated. The main sensitivity is to the speed of melt flow and the mantle upwelling rate (which depends on the spreading rate). Both of these effects are contained within the parameter $\mathcal{Q}$. Our standard choice of $\mathcal{Q}$ results in melt flow that is comparable in magnitude to previous studies \citep{crowley2015glacial,burley2015variations} but slower than has been inferred for melt velocity based on the Iceland post-glacial melt pulse \citep{Maclennan2002,Swindles2017,Prin2019}). We explore model sensitivity to the choice of $\mathcal{Q}$ in figure~\ref{fig:admittance_wet} and the Supplementary Information (section S1).

The melting column (Figure~\ref{fig:base}) can be divided into three regions: `wet', `transitional' and `dry', as labelled on the figure. In the wet region, occupying roughly the bottom half of the column, a small amount of carbon-rich melts are generated. The porosity and hence the liquid flux remain very small. There is almost no melt segregation. In the dry region, occupying roughly the top half of the column, the degree of melting increases. The liquid flux increases and the solid flux decreases. The transport of carbon becomes dominated by the liquid phase. Around the depth of the carbon-free solidus, there is a transitional region separating the wet and dry regions. We discuss the steady-state behaviour and choice of parameters further in \ref{app:steady}. 

Two important scales emerge from the steady solution: the maximum porosity at the top of the column
\begin{equation}
    \phi_\mathrm{max} \sim \left(\frac{F_\mathrm{max} }{\mathcal{Q}}\right)^{1/n} = \left(\frac{F_\mathrm{max} \mu W_0 }{k \Delta \rho g}\right)^{1/n},
\end{equation}
and the liquid velocity at the top of the column, which, in dimensional units, is given by
\begin{equation}
    w_0 \sim W_0 \frac{F_\mathrm{max} }{\phi_\mathrm{max} }.
\end{equation}
Thus, if the melt flux parameter $\mathcal{Q}$ is large, melt extraction is efficient and so the maximum porosity is small and the melt velocity is large.
The dimensional melt transport time $t_\mathrm{dry}$ across the dry-melting region is approximately given by 
\begin{equation} \label{eq:t_melt}
    t_\mathrm{dry} \sim n \frac{H_\mathrm{dry}}{w_0},
\end{equation}
where the factor of $n$ arises from the $z$-dependence of the velocity. 

The variations in sea level associated with glacial cycles are on the scale of 100~m, while the depth of the melting region is tens of kilometres. Furthermore, the density ratio $\rho_w/\rho \approx 0.3 $. These considerations imply that the variations $\delta$ in the depth of first partial melts associated with sea-level change are small, according to equation~\eqref{eq:delta_def}. Therefore, we assume that the time-dependent fluctuations of all quantities are relatively small and linearize the governing equations about the steady state. We further decompose the time-dependent fluctuations into harmonics of dimensionless frequency $\omega$. Thus 
\begin{equation}
    \delta = \delta_0 \mathrm{e}^{i\omega t},
\end{equation}
where $\delta_0 \ll 1 $ is the maximum fluctuation. This is appropriate for either periodic forcing of frequency $\omega$ or by taking the Fourier transform of a record of sea level (in the latter case, $\delta_0$ is a function of $\omega$). 
The dimensionless sea level satisfies
\begin{equation}
    S=S_0 + \frac{\Delta S}{H} \mathrm{e}^{i\omega t},
\end{equation}
where $\Delta S$ is the maximum dimensional sea-level fluctuation. If we define $\Delta \delta\equiv 2\delta_0$, then $\Delta S=H \Delta \delta (\rho/\rho_w)$. 

We decompose the porosity and bulk carbon concentration into steady parts (denoted with an overline) and time-dependent parts (denoted with a prime), 
\refstepcounter{equation}
\[
    \label{eq:decomposition-defintion}
    \phi = \overline{\phi}(z) + \phi'(z,t), \qquad  
    c = \overline{c}(z) + c'(z,t),
    \eqno{(\theequation{\mathit{a},\mathit{b}})} 
\]
where $\phi'(z,t) = \delta_0\mathrm{e}^{i\omega t} \hat{\phi}(z)$ and $c'(z,t) =\delta_0 \mathrm{e}^{i\omega t} \hat{c}(z)$. We derive equations governing the fluctuations of porosity and carbon concentration in Appendix~\ref{sec:perturbed}. These are used to deduce the contribution of sea-level variations to the melt and carbon fluxes. Equivalent notation is used for the decomposition of the fluxes into steady and fluctuating parts. The fluctuating part of the carbon flux has contributions from both the fluctuation of melt flux and carbon concentration:
\begin{equation} \label{eq:carbon_flux}
    Q'_c = \frac{ Q' \overline{c} + \overline{Q} c'}{\mathcal{M} D_c}.
\end{equation}

Software to reproduce the calculations and the results shown in the figures in the remainder of this paper is available \citep{code}.

\begin{table} 
\centering
\caption{Physical parameters and corresponding dimensionless model parameters. Values as used in calculations unless otherwise stated. See~Appendix~\ref{app:steady} for justification. } \label{tab:parameters}
\begin{tabular}{c c c l} 
\hline 
Parameters & Value & Unit & Description \\
\hline 
$\rho_w$					& 1000			& kg/m$^{3}$ & Sea-water density \\ 
$\rho$						& 3300			& kg/m$^{3}$ & Mantle density \\
$W_0$  					& 2 				& cm/yr 			& Mantle upwelling velocity \\ 
$H$	 						& 130 			& km 				& Height of melting column \\
$H_{\text{dry}}$	 	& 65 			& km 				& Height of dry melting column \\
$\Delta S$	 	& 0.1 			& km 				& Peak-to-trough amplitude of sea-level fluctuation \\
$F_{\text{max}}$ 		& 0.2 			& 						& Maximum degree of melting \\
$D_c$   					& $10^{-4}$ & 						& Partition coefficient of carbon \\ 
$n$       					& 2  				& 						& Exponent in permeability-porosity relationship \\
$\mathcal{Q}$  		 	& $10^{5}$ & & Liquid flux scale	\\ 
$\Gamma^*$  		 	& $0.4$ & &	Melting rate scale\\ 
$\mathcal{M}$  		& $0.2$ & & Effect of carbon on the mantle solidus scale	\\ 
$\phi_\mathrm{max}$  		& $0.0014$ & & Maximum steady-state porosity at top of column	\\ 
$w_0$  		& $2.8$ & m/yr & Maximum melt velocity at top of column	\\ 
$t_\mathrm{dry}$  		& $46$ & kyr & Melt transport time across dry melting region	\\ 
$\alpha$    & 30   & $^\circ$   & Dip of decompaction channel \\
\hline
\end{tabular}
\end{table}

\subsection{Pseudo-two-dimensional model of melt focusing} \label{sec:2d-description}

The melting region beneath a mid-ocean ridge is not columnar; rather it is a volume that encloses upwelling, melting mantle \citep{Forsyth1998MELT}. In a vertical plane normal to the ridge axis, the shape of the melting region can be approximated as triangular \citep{langmuir1992petrological}. Magma produced off-axis is focussed along a decompaction channel at the base of the lithosphere toward the ridge axis \citep{sparks1991melt}.

Figure~\ref{fig:sketch} illustrates our pseudo-two-dimensional model (see  Appendix~\ref{sec:focusing_model} for full details).  We assume that the melting region comprises an array of independent columns that deliver magma into a decompaction channel \citep{sparks1991melt}. The decompaction channel transports both the mean flux and variations.  Following previous work, we consider two simple assumptions for this transport: instantaneous \citep{burley2015variations} and finite-rate \citep{crowley2015glacial}.  Magma and carbon flux variations are delivered to the ridge according to an integral over columns from the axis out to some maximum focusing distance (equation~\eqref{eq:focusing-model}). In the case of finite-rate focusing, the transport time causes a phase-delay $\tau$ that increases with distance to the ridge $x$.

\section{Results} \label{sec:Results}
Variation in sea level causes variation in pressure and hence variation in (\textit{i}) the onset depth of melting and (\textit{ii}) the melting rate throughout the column. We refer to (\textit{i}) as the `basal-flux mechanism' and to (\textit{ii}) as the `internal-melting mechanism'. These mechanisms, in turn, drive variation in the melt and carbon flux.

\subsection{Example of fluctuations due to sea-level changes} \label{sec:results_example}

\begin{figure}
    \centering 
    \includegraphics[width =0.95\linewidth]{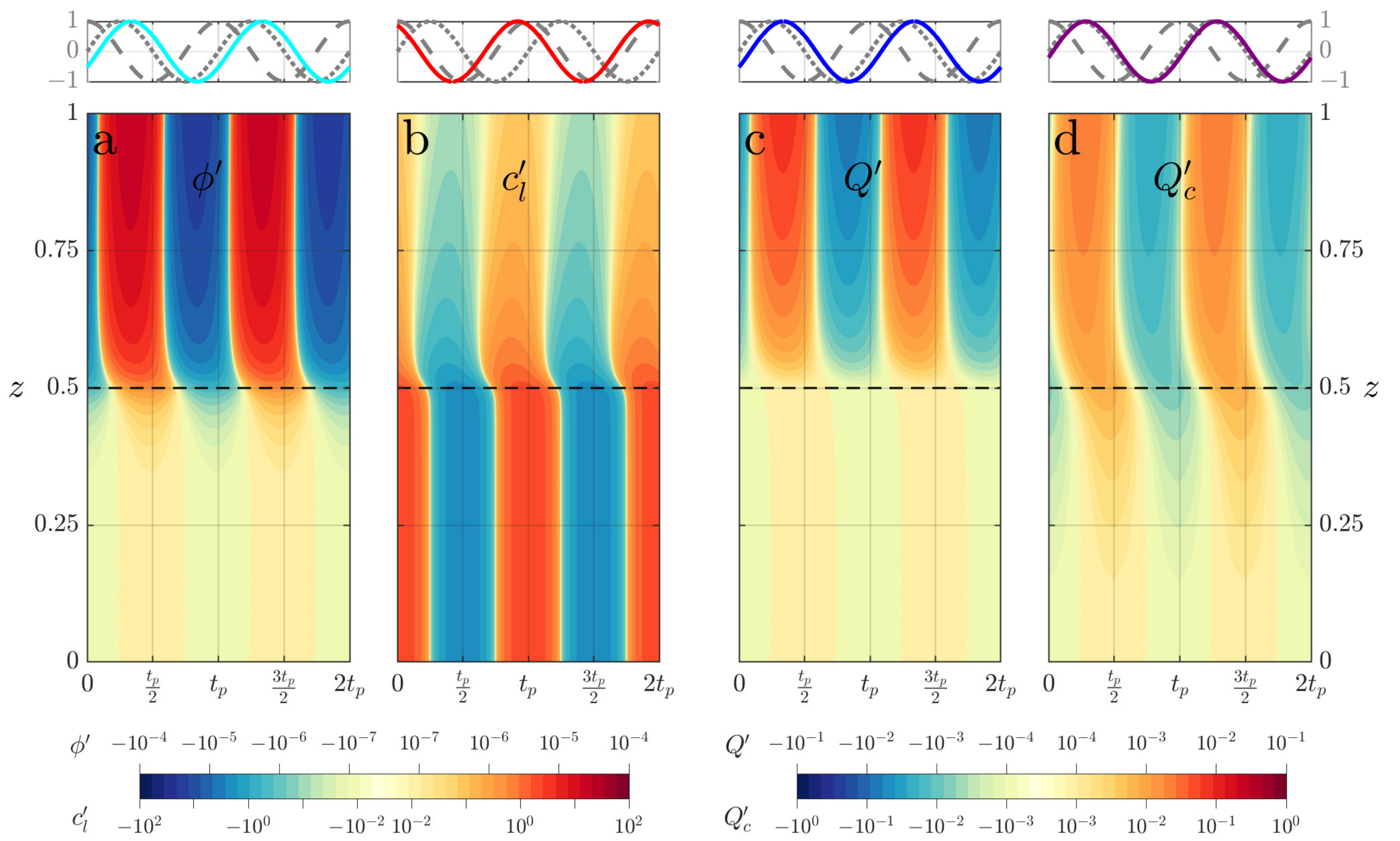}
    \caption{Time-dependent fluctuations caused by sea-level cycles with a period of 100~kyr and a peak-to-trough amplitude of 100~m. From left to right: (a) porosity $\phi'$, (b) carbon concentration in the melt $c_l'$, (c) melt flux $Q'$, and (d) carbon flux $Q_c'$. The wet-to-dry transition is indicated by the dashed line at $z=0.5$. The colour bars report quantities non-dimensionalized as described in the text. In the first row, we display the time-evolution of the four fields at $z=1$ (solid lines), the sea-level variation $S$ (dashed line) and the rate of sea-level decrease $-\dot{S}$ (dotted line). All the curves in the first row are normalized to have the same amplitude to facilitate comparison of the phase.}
    \label{fig:PS_example}
\end{figure}

Figure~\ref{fig:PS_example} shows the response of porosity, carbon concentration, melt flux and carbon flux to a sea-level cycle that has a peak-to-trough magnitude of 100~m and a period of 100~kyr. While these numbers are chosen for illustration, they roughly correspond to the sea-level variation experienced in the late Pleistocene (past 800~kyr).

We first discuss the coupled evolution of the porosity and carbon concentration, since these are the primary fields. We discuss the evolution working from the bottom of the melting column to the top, following the direction of the steady-state liquid and solid flow. 

Figure~\ref{fig:PS_example}a shows the porosity fluctuation. In the wet-melting region, near the bottom of the melting column, the basal-flux mechanism is significant. Here the steady-state porosity increases with height (figure~\ref{fig:base}a) and hence, in order to maintain zero porosity at the onset of melting $z=\delta(t)$, a positive sea-level fluctuation leads to a negative porosity fluctuation. Conversely, the steady-state carbon concentration  decreases with height (figure~\ref{fig:base}b). So a positive sea-level fluctuation leads to a positive carbon fluctuation (figure~\ref{fig:PS_example}b). Hence the porosity is in antiphase with sea level but the carbon concentration is in phase. The magnitude of the porosity variation is relatively small because the steady-state porosity is also very small. By contrast, the magnitude of the carbon variation is more significant because of the sharper variation in the steady carbon concentration. The internal-melting mechanism causes little change to $\phi$ and $c_l$ because there is no melt segregation in the wet-melting region. Thus there is an almost perfect balance between the melting-rate fluctuation caused by decompression melting and that caused by the carbon concentration fluctuation. Carbon is buffering the system to counteract the extra internal decompression melting caused by sea-level variation. In terms of the melting rate given by equation~\eqref{eq:melting_rate}, \mbox{$\Gamma_P \sim -\Gamma_c$}. 

In the transitional region, around the depth of the dry solidus, the fluctuations of porosity and carbon concentration change significantly. It is in this zone that the melt segregation becomes important, since this is where the steady porosity and melt flux start to increase significantly. Melt segregation breaks the buffering capacity of carbon described previously \mbox{$(\Gamma_p \nsim -\Gamma_c)$} because $\Gamma_c$ is reduced in magnitude. Indeed, segregation leads to a decrease in the magnitude of the carbon fluctuation. There is also a shift to a slightly positive phase (N.B., throughout the manuscript, a `positive phase' is achieved when the peak of a fluctuation shifts to earlier times, as in the transition region in figure~\ref{fig:PS_example}b). Furthermore, the porosity fluctuation increases with depth and undergoes a phase shift, which we explore below.

In the dry-melting region, near the top of the column, the dominant contributions to porosity evolution come from internal decompression melting $\Gamma_p$ and fluctuations in the upward transport of melt.  The resultant phase observed is intermediate between the phase associated with the basal flux ($-S$) and that associated with internal melting ($-\dot{S}$, where a dot represents a time derivative). Concurrently, the carbon fluctuation continues to decrease in magnitude, in the manner outlined above. 

The melt flux fluctuation (figure~\ref{fig:PS_example}c) has the same phase as the porosity fluctuation. The amplitude of the flux increases more rapidly with height than porosity because the steady-state flux also increases. The carbon flux fluctuation (figure~\ref{fig:PS_example}d) has contributions from both the melt flux and carbon concentration fluctuations, which must be weighted by the steady state as given by equation~\eqref{eq:carbon_flux}. The former contribution is larger, so the carbon flux largely follows the melt flux fluctuation. The carbon concentration fluctuation has the opposite phase so, in part, offsets the melt flux fluctuation and slightly advances the phase, so that the peak in carbon flux occurs slightly earlier than the peak in melt flux. 

\subsection{Effect of the period of sea-level fluctuations} \label{sec:results_period}
\begin{figure}
    \centering 
    \includegraphics[width =0.8\linewidth]{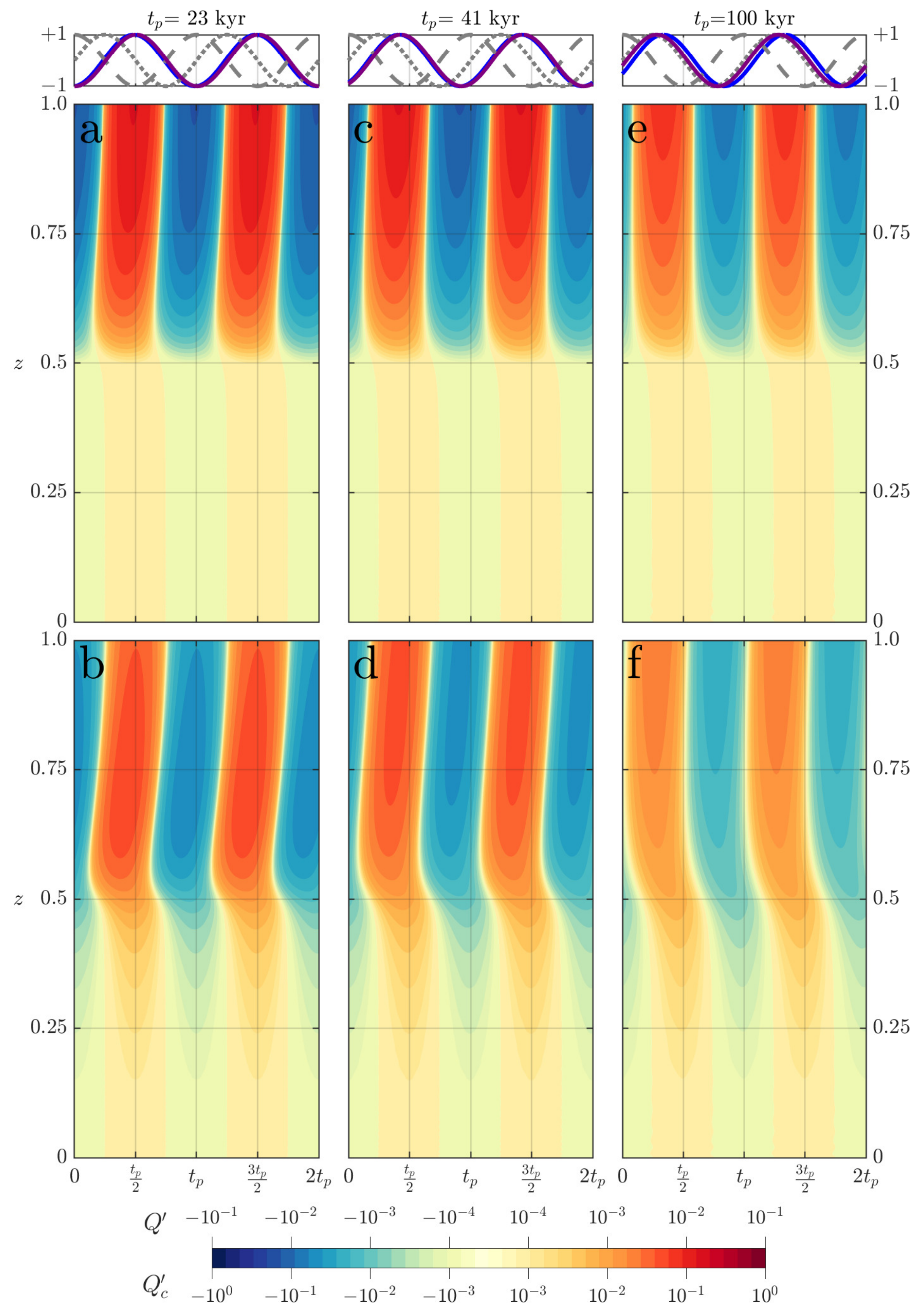}
    \caption{Fluctuations of melt flux (2nd row, panels a,c,e) and carbon flux (3rd row, panels b,d,f) at forcing periods of 23~kyr (1st column, panels a,b), 41~kyr (2nd column, panels c,d) and 100~kyr (3rd column, panels e,f). On the first row, the blue lines and purple lines correspond to the time-evolution of melt and carbon fluxes, respectively, at $z=1$. For other legend details see Figure~\ref{fig:PS_example}. Note that panels e and f are identical to figure~\ref{fig:PS_example}c and d; they are repeated here to facilitate comparison.}
    \label{fig:PS_Wet_fluxes}
\end{figure}

These flux variations are sensitive to the period of forcing. Figure~\ref{fig:PS_Wet_fluxes} shows the melt and carbon flux fluctuations at three different periods, reflecting the dominant periods of sea-level variations in the late Pleistocene. We focus our discussion on the behaviour at the top of the melting column. At very short periods (e.g.,~23~kyr), the melt and carbon flux fluctuations are in antiphase with sea level. With increasing forcing period, there is a switch in behaviour. For a sufficiently long period (e.g.,~100~kyr), the system behaves as described in section~\ref{sec:results_example}, with fluxes proportional to the rate of decrease in sea level. This results in a phase advance of about a quarter of a period as the forcing period increases.

The behaviour of the system with a short forcing period (high frequency) can be approximated using an analytical method. A detailed calculation is presented in Appendix~\ref{sec:large_omega}. Here, we describe the main physical ideas and insights of this analysis. At high forcing frequency, melt segregation during one cycle is minimal and the porosity and carbon concentration respond almost instantaneously throughout the melting column to the pressure change induced by the change in sea level. 

This behaviour changes when the forcing period is proportional to the melt travel time across the dry melting region $t_\mathrm{dry}$ given by equation~\eqref{eq:t_melt}, and occurs at a dimensional critical period $t_p^*$ given by
\begin{equation} \label{eq:critical_tp}
    t_p^* = C \frac{1}{n(n-1)} t_\mathrm{dry} = C \frac{1}{n-1} \frac{H_\mathrm{dry}}{w_0},
\end{equation}
where $C \approx 1$ is a dimensionless prefactor (roughly corresponding to a 20\% phase shift). For the parameters given in table~\ref{tab:parameters}, the critical period is about 23~kyr, consistent with figure~\ref{fig:admittance_wet}b described below. To a good approximation, the critical period does not depend on the solidus-depressing thermodynamic effect of carbon. Similar arguments can be applied to the evolution of carbon, in which case dilution by melt transport determines the critical period given by equation~\eqref{eq:critical_tp_app_Qc} in Appendix~\ref{sec:large_omega}.

\subsection{Admittance and lag}

\begin{figure}
    \centering 
    \includegraphics[width =0.75\linewidth]{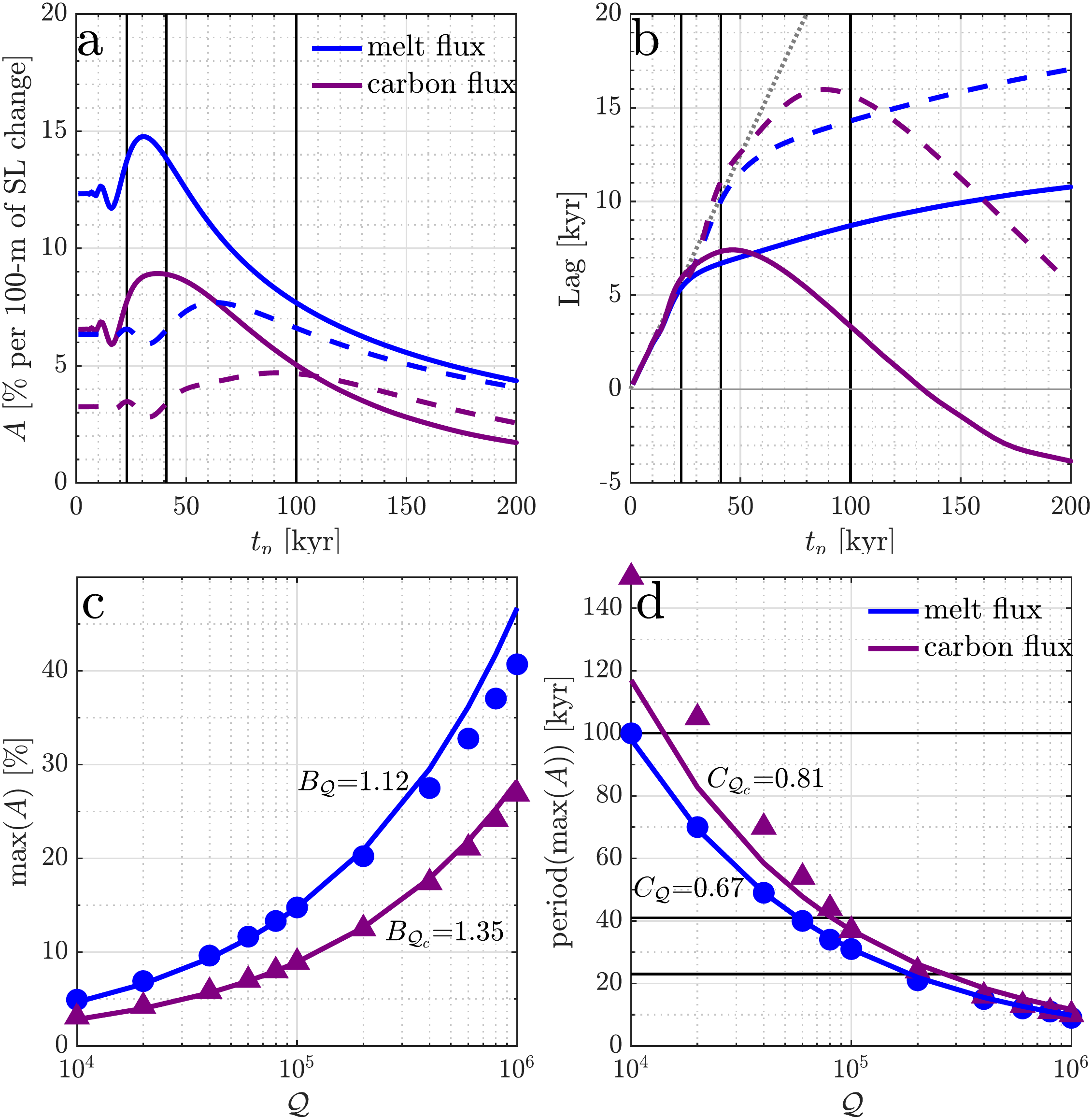}
    \caption{Sensitivity of fluctuations to forcing period and melt velocity scale. First row: Admittance (a) and lag (b) of melt flux (blue line) and carbon flux (purple line) with dimensional forcing period $t_p$ at $\mathcal{Q}=10^{5}$ (solid lines) and $\mathcal{Q}=0.25\times10^{5}$ (dashed line). To help the analysis of lags, we indicate $t_p/4$ by a grey dotted-line. 
    Second row: Value of the maximum admittance (c) and forcing period $t_p^\mathrm{max}$ at which this maximum occurs (d) as a function of $\mathcal{Q}$. Symbols represent the calculations from our models and solid lines are analytical estimations (see the main text for explanations and Appendix~\ref{sec:large_omega} for details). The prefactors $B_Q,B_{Q_c}$ and $C_Q,C_{Q_c}$ are chosen so that the analytical estimates fit our model calculations at the reference value of $\mathcal{Q}$. The main Pleistocene periods (23, 41, and 100~kyr) are indicated in sub-panels a,b and d by the solid black lines.}
    \label{fig:admittance_wet}
\end{figure}

Figure~\ref{fig:admittance_wet} summarizes the effect of the period of sea-level forcing in terms of the admittance and lag of melt and carbon fluxes. Here the admittance (panel~a) is defined as the peak-to-trough magnitude of the time-dependent part of the flux at the top of the melting column normalized by the steady-state flux. This quantity (sometimes called `relative admittance') is proportional to the amplitude of the sea-level cycle. Therefore, we report admittance as a percentage per 100~m peak-to-trough sea-level fluctuation. We define the lag (panel~b) as the difference between the time of the peak flux and the time of peak rate of decrease in sea level $-\dot{S}$. This choice of baseline, while somewhat arbitrary, is motivated by the fact that it corresponds to a maximum in $\Gamma_P$, i.e., to the peak rate of generation of extra partial melts due to sea level. 

The calculated lag at small forcing periods is approximately proportional to $t_p/4$ because the fluxes are antiphase with sea level, rather than in phase with the rate of decrease in sea level. At periods longer than the critical period of equation~\eqref{eq:critical_tp}, the phase approaches that corresponding to the rate of decrease in sea level. Thus, although the calculated lag of melt flux continues to increase with increasing period (reflecting the finite melt transport time), it does so at a much smaller rate. 

At long periods, the admittance decreases with period because longer period corresponds to slower sea-level change. The melt flux admittance $A_Q$ can be estimated \citep{lund2011does} by comparing the effect of sea-level variation on pressure to the static pressure and hence the relative melting rate ($\Gamma_P/\Gamma_0$). This gives
\begin{equation} \label{eq:estimate_A_Q_long_tp}
    A_Q \sim \frac{\Delta S}{W_0}\frac{\rho_w}{\rho} \frac{2 \pi}{t_p},
\end{equation}
which is a good estimate at periods longer than about 100~kyr [see Supplementary Information, section S2].

Conversely, at short periods, the admittance tends toward a constant that is independent of period.  The magnitude of this constant can be estimated analytically, as done in Appendix~\ref{sec:large_omega} where we derive the following approximations for admittance of porosity $A_\phi$, melt flux $A_Q$ and carbon flux $A_{Q_c}$,
\begin{linenomath*} \begin{subequations}
\begin{align}
    & A_\phi  \sim \frac{\Delta S}{W_0} \frac{\rho_w}{\rho} \frac{w_0}{H_\mathrm{dry}} , \label{eq:estimate_A_phi} \\
    & A_Q  \sim n \frac{\Delta S}{W_0} \frac{\rho_w}{\rho} \frac{w_0}{H_\mathrm{dry}}, \label{eq:estimate_A_Q} \\
    & A_{Q_c} \sim (n-1) \frac{\Delta S}{W_0} \frac{\rho_w}{\rho} \frac{w_0}{H_\mathrm{dry}}, \label{eq:estimate_A_Q_c}
\end{align}
\end{subequations} \end{linenomath*}
where $\Delta S$ is the dimensional magnitude of the sea-level fluctuation and $w_0$ is the dimensional melt velocity at the top of the column. The physical meaning of these expressions can be interpreted as follows, making the approximation of negligible melt segregation on the timescale of one period of sea-level variation. The fluctuating part of the porosity is equal to the pressure change associated with sea level multiplied by the productivity. The steady-state melt flux (which is equal to the porosity multiplied by the melt velocity) is equal to the pressure change across the dry melting region multiplied by the productivity and the mantle upwelling rate. This allows us to estimate the steady-state part of the porosity. The combination of the estimates for the steady and fluctuating parts of the porosity shows that the admittance is equal to the ratio of melt to mantle velocity multiplied by the ratio of pressure change associated with sea level to that across the dry melting region, giving equation~\eqref{eq:estimate_A_phi}. The admittance of melt flux is a factor of $n$ larger than that of porosity because melt flux increases with porosity as a power law with exponent $n$, so $A_Q = n A_\phi$, giving equation~\eqref{eq:estimate_A_Q}. The fluctuation of carbon concentration is the opposite to that of porosity; this is required to keep the bulk concentration constant, so $c' \overline{\phi} = -\overline{c} \phi'$. Physically, an increase in porosity dilutes the carbon concentration in the melt. Finally, by combining this with the carbon flux fluctuation given by equation~\eqref{eq:carbon_flux}, we obtain equation~\eqref{eq:estimate_A_Q_c}. 

Figure \ref{fig:admittance_wet}(c,d) shows that this theory can be applied to obtain simple estimates of the maximum admittance and forcing period at which that maximum occurs. The maximum admittance of melt flux occurs at a forcing period $t_p^\mathrm{max}$ that is proportional to the melt transport time across the dry-melting region $t_\mathrm{dry}$, as previously suggested by \citet{crowley2015glacial}. In particular, we write
\begin{equation} \label{eq:tp_max}
    t_p^\mathrm{max} \sim C_{Q}\, t_\mathrm{dry}.
\end{equation}
where $C_{Q}=0.67$ is a prefactor chosen so that the analytical estimations fit our model calculations at the reference value of $\mathcal{Q}$. This period is intermediate between the long- and short-period limits.
The magnitude of the maximum admittance of melt flux is a factor of about $B_Q=1.12$, i.e., 12\% greater than the approximate formula~\eqref{eq:estimate_A_Q} applicable for small forcing period.  Physically, this maximum occurs when the porosity fluctuations caused by melting are positively reinforced by fluctuations of the upward transport of melt (these contributions reinforce each other at intermediate $t_p$). 

The same approach can be applied to carbon fluxes. We use the same notation for the prefactors, except replacing subscript $_Q$ (melt flux) with subscript $_{Q_c}$ (carbon flux) and using equation~\eqref{eq:estimate_A_Q_c}, which is the estimate of carbon flux admittance at small forcing period.

These results are sensitive to the fluid dynamical properties of the system. Figure~\ref{fig:admittance_wet}(a,b) shows the effect of reducing melt flux parameter $\mathcal{Q}$ by a factor of 4. This could correspond, for example, to a reduction of permeability, an increase in melt viscosity or an increase in mantle upwelling rate. This reduces liquid velocities and hence increases the melt transport time. The admittance is reduced. At small periods, $A_Q$ is reduced by a factor of 2 ($=4^{1/n}$, since $n=2$, see equation~\eqref{eq:admittance_melt_approx}). At large periods, the effect is much more modest. Indeed, reducing the permeability can slightly increase the carbon flux fluctuation at sufficiently long periods. The lag is typically (but not always) increased by increasing the melt transport time (smaller $\mathcal{Q}$) and the critical period $t_p^*$ increases, consistent with equation~\eqref{eq:critical_tp}. The behaviour of the carbon flux is complicated because it is affected by two contributions: one from the melt flux fluctuation and the other from the carbon concentration fluctuation, as shown in equation~\eqref{eq:admittance_carbon_approx}. The latter tends to be in phase with $S$ and so peaks earlier in the cycle, which leads to the negative lags with respect to $-\dot{S}$ at high forcing periods (figure~\ref{fig:admittance_wet}b). Carbon concentration fluctuations are also responsible for the non-monotonic sensitivity of admittance to permeability.

\section{Discussion} \label{sec:Discussion}

This study builds on previous work by incorporating fluctuations in the melting rate throughout the melting column (`internal melting') and also by considering the thermodynamic effect of carbon on melting. \citet{burley2015variations} considered only fluctuations introduced by variation in the melting-onset depth with sea level (`basal flux'). \citet{crowley2015glacial} considered only internal melting and calculated only melt fluxes. They did not calculate the concentration of carbon, nor consider its thermodynamic effect. 

The simplifying assumptions made in previous studies can be tested within our framework. First, the thermodynamic effect of carbon can be assessed using a `dry-melting model' ($\mathcal{M}= 0$). Second, the importance of internal melting can be assessed by excluding it from the equations governing the fluctuations (in addition to $\mathcal{M}= 0$, we force $\Gamma_P=0$; see Appendix~\ref{sec:perturbed}). We label this a `basal-flux model' because it includes only this mechanism.

\begin{figure}
    \centering 
    \includegraphics[width =0.8\linewidth]{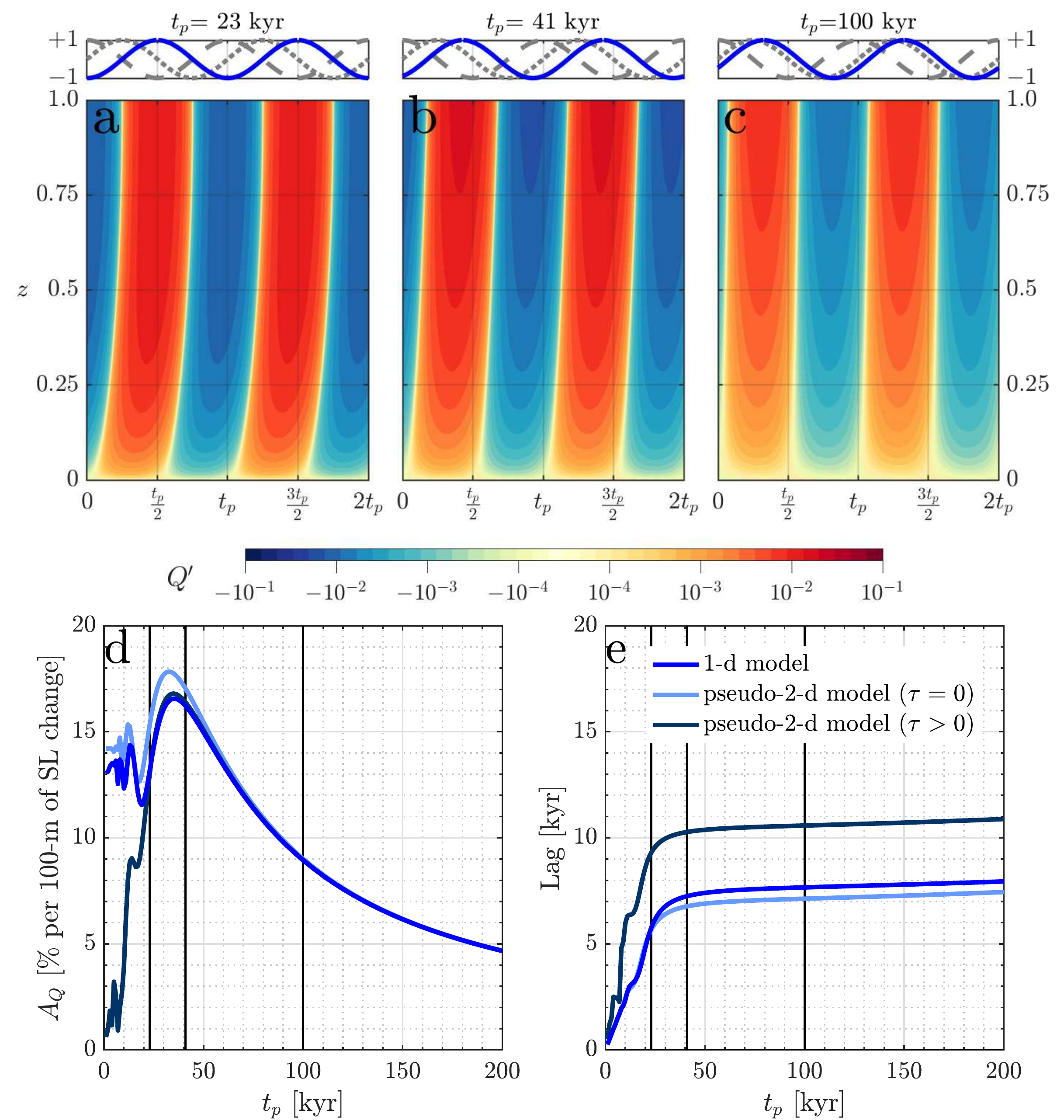}
    \caption{Dry-melting models ($\mathcal{M}=0$). Fluctuations of melt flux at forcing periods of 23~kyr~(a), 41~kyr~(b) and 100~kyr~(c). On the first row, the blue lines correspond to the time-evolution of melt flux at $z=1$. For other legend details see figure~\ref{fig:PS_example}. We also show the admittance~(d) and lag~(e) for both the 1d model and pseudo-2d models. The pseudo-2d model with $\tau > 0$ is equivalent to the model of \citet{crowley2015glacial}}
    \label{fig:DryMelt}
\end{figure}

Figure~\ref{fig:DryMelt} shows that there are minimal differences between our dry- and wet-melting models near the surface. This is because carbon has only a minor effect on melting outside the wet-melting region. Fluctuations in melt flux and the melt-flux admittance (blue line figure~\ref{fig:DryMelt}d) behave similarly in both cases. In particular, $A_Q$ approaches a constant admittance at small forcing periods. This finding contrasts with \citet{crowley2015glacial}, so we test whether this difference arises from our 1d simplification by creating pseudo-2d models using the same methodology (see section~\ref{sec:2d-description} and Appendix~\ref{sec:focusing_model}).
The admittance of the pseudo-2d model with instantaneous focusing (light blue, figure~\ref{fig:DryMelt}d) is similar to that of the 1d model. In contrast, when assuming a finite-focusing rate (dark blue, figure~\ref{fig:DryMelt}d, we predict sharp decrease in the melt-flux admittance at small period. The latter case is consistent with the model of \citet{crowley2015glacial}. However, at forcing periods of Milankovitch cycles, there are minimal differences between 2d and 1d models.  In addition to 2d effects, storage in crustal magma chambers and turbulent mixing of carbon from the MOR to the surface ocean and atmosphere would attenuate admittance at timescales less than about 1~kyr. 

\begin{figure}
    \centering 
    \includegraphics[width =0.8\linewidth]{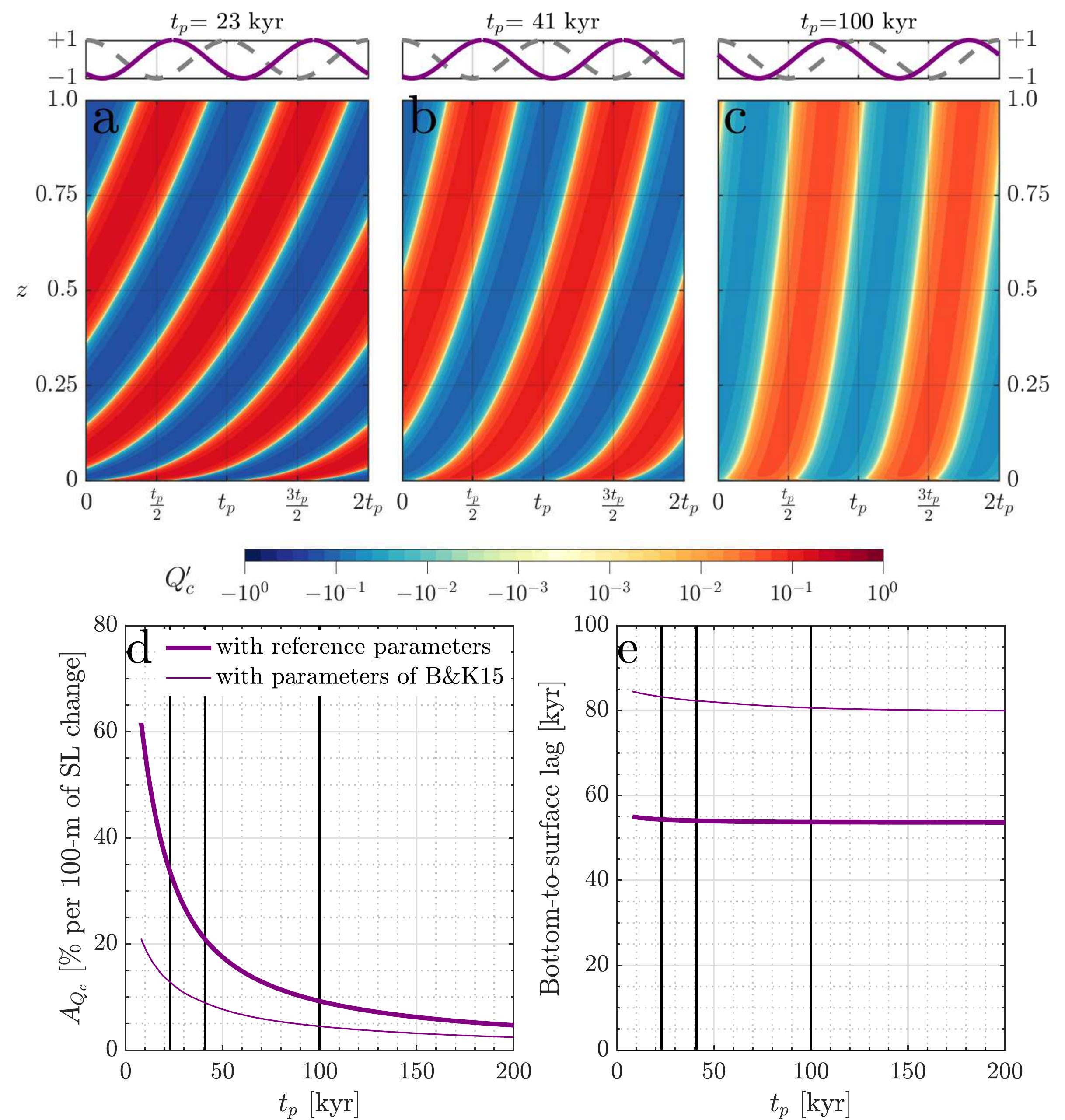}
    \caption{Basal-flux models ($\mathcal{M}=0,\,\Gamma_P=0$) akin to the models of \citet{burley2015variations}. Fluctuations of carbon flux at forcing periods of 23~kyr~(a), 41~kyr~(b) and 100~kyr~(c). On the first row, the purple lines correspond to the time-evolution of melt flux at $z=1$. For other legend details see figure~\ref{fig:PS_example}. We also show the admittance~(d) and bottom-to-surface lag~(e). For comparison we have computed both quantities for a model with same parameters as \citet{burley2015variations}. Also, for consistency with \citet{burley2015variations}, we report lag as the time required for a maximum (or minimum) in carbon flux to travel from the base to the top of the column. Note that the maximum flux near the base occurs around $t=t_p/4$, corresponding to a peak in $-\dot{S}$ (as argued by \citet{burley2015variations}). This is most clear in panel (c). To account for this, we add $t_p/4$ to the lag to obtain an effective lag. This effective lag is almost independent of forcing period and reflects the bottom-to-surface melt segregation time, consistent with \citet{burley2015variations}.}
    \label{fig:DryBasal}
\end{figure}

Figure~\ref{fig:DryBasal} shows our basal-flux model, which we now compare to figures 6a and 7c in \citet{burley2015variations}. 
In both cases, the admittance decreases with forcing period and the lag is insensitive to forcing period. Using the same parameters as \citet{burley2015variations} (thin line), we find a good agreement between our models. Thus we regard our basal-flux model as a one-dimensional representation of \citet{burley2015variations}. In our model with reference parameters (thick line; larger permeability than in \cite{burley2015variations}), the admittance is increased and the lag is diminished. At a forcing period of 100 kyr, the admittance is two times higher and the lag is one third lower than in the models with values of \citet{burley2015variations}. Next, we compare the basal-flux model, and thus \citet{burley2015variations}, with the full wet-melting model (figures~\ref{fig:PS_Wet_fluxes} and \ref{fig:admittance_wet}). The basal-flux model gives a slightly greater admittance of carbon flux at the dominant forcing periods and a different admittance structure at small periods. There are also profound differences in the timing of the surface fluxes because, in the basal-flux model, fluctuations created at the base must travel to the surface, which occurs approximately over the melt travel time. By contrast, accounting for internal melting means that much of the effect of sea-level fluctuation is generated closer to the surface (this is true independent of volatile content). Thus our more general models that incorporate both basal-flux and internal-melting mechanisms differ markedly from \citet{burley2015variations} in terms of the lag predicted.

\begin{figure}
    \centering 
    \includegraphics[width =0.85\linewidth]{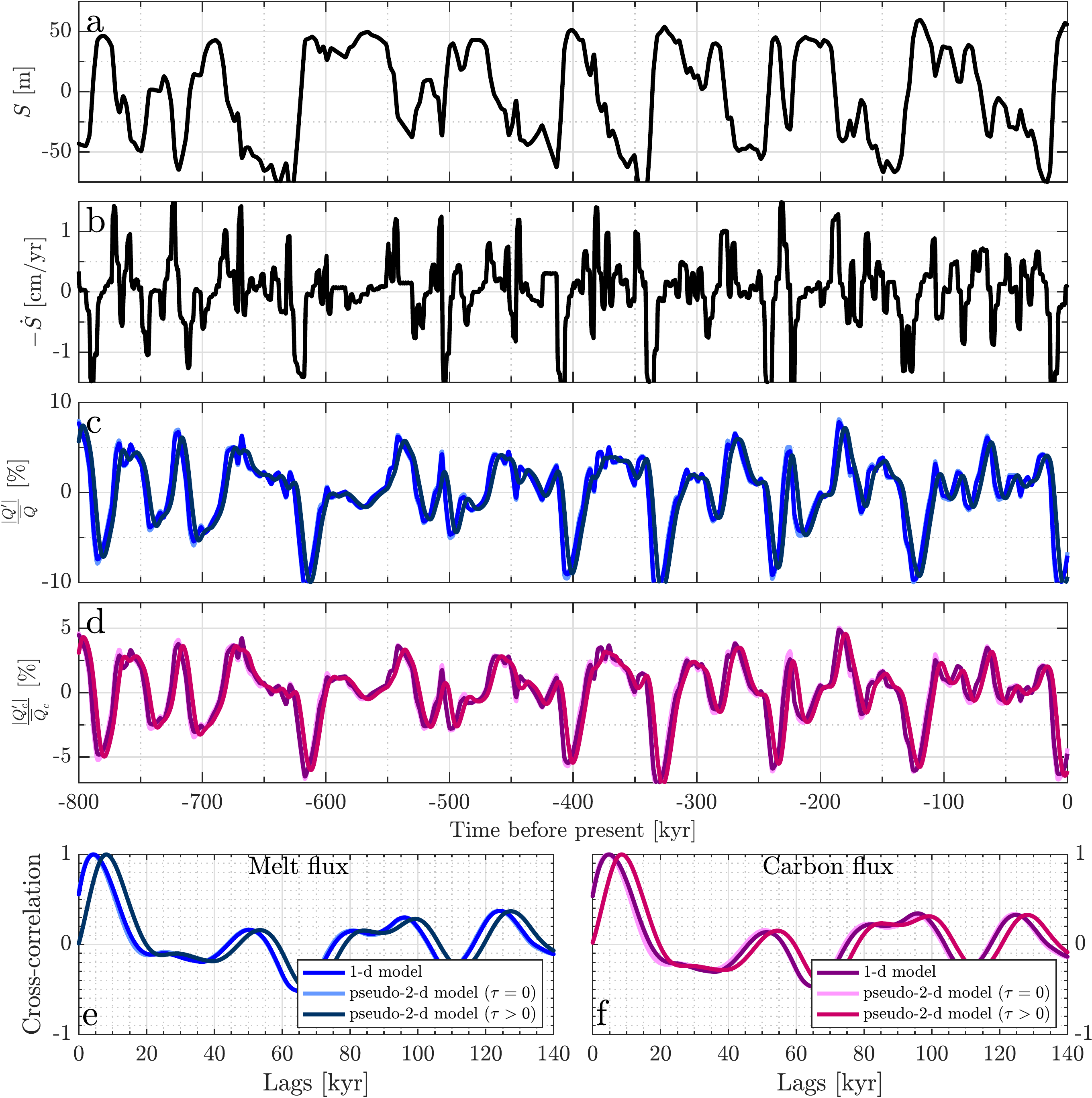}
    \caption{Predictions of the effects of sea-level variations over the past 800~kyr on melt and carbon fluxes. (a) Time series of reconstructed global sea level from \citet{siddall2010changes}. (b) Rate of decrease of the reconstructed sea level, which are useful when comparing to the results below. The remaining rows show the results of our calculations forced by the sea-level record, using 1d and pseudo-2d models. Variations in melt flux (c) and carbon flux (d) are expressed as a percentage of the steady-state values at the surface. Cross-correlation lag of the rate of sea-level decrease ($-\dot{S}$) with melt (e) and carbon fluxes (f). Cross-correlation values have been normalized to their maximum.}
    \label{fig:Column_TimeSeriesSL}
\end{figure}

\begin{figure}
    \centering 
    \includegraphics[width =0.75\linewidth]{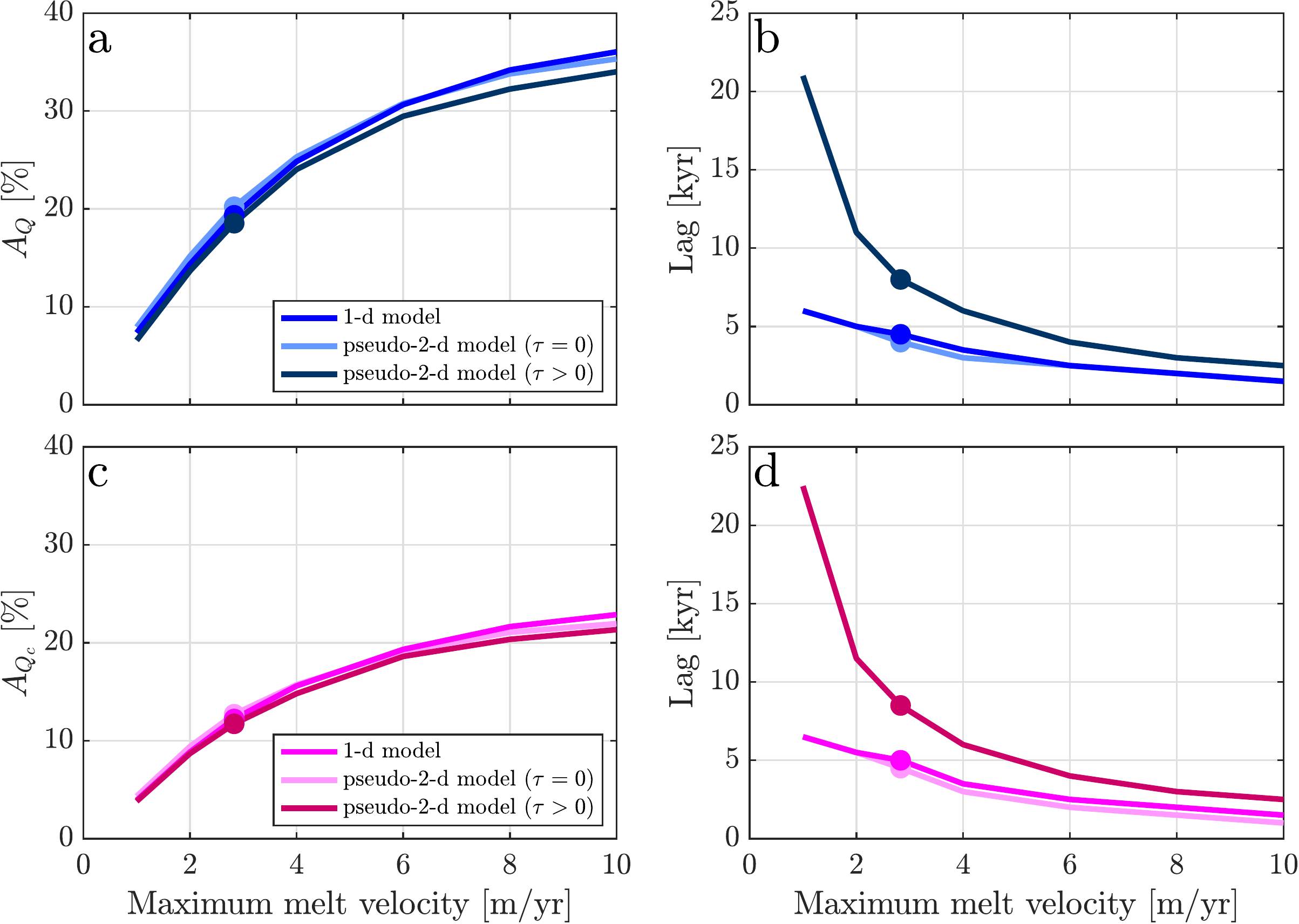}
    \caption{Admittance and maximal cross-correlation lag of melt and carbon flux in models forced with reconstructed global sea level as a function of the dimensional maximum steady-state melt velocity (fig.~\ref{fig:Column_TimeSeriesSL}). Results are sensitive to the ratio of this velocity to the mantle upwelling velocity and hence to spreading rate. Therefore, increasing the maximum melt velocity by a factor of 2 is equivalent to halving the spreading rate. Dots indicate values calculated with the reference parameters of table \ref{tab:parameters} and correspond to figure \ref{fig:Column_TimeSeriesSL}.}
    \label{fig:Column_SLrecord_varQ}
\end{figure}

Finally, we assess the implications of our calculations for carbon fluxes from mid-ocean ridges over the Pleistocene. Figure~\ref{fig:Column_TimeSeriesSL} shows our predictions based on the reconstructed sea-level variation over the past 800~kyr. Our calculations indicate that sea-level fluctuations have driven substantial variation in melt and carbon fluxes from the mid-ocean ridge system.  The melt and carbon fluxes depart from the mean, steady-state values with a total range of about 20\% and 13\% respectively, and are therefore potentially significant contributors to variation in crustal thickness and variation in magmatic carbon fluxes to the ocean/atmosphere. Figure~\ref{fig:Column_SLrecord_varQ}a,c shows that these estimates are particularly sensitive to the melt transport (reported as a function of steady-state melt velocity at the top of column) but not strongly influenced by 2d effects.

The melt and carbon fluxes in figure~\ref{fig:Column_TimeSeriesSL} are most closely related to the reconstructed rate of decrease in sea level $-\dot{S}$. Variations of this rate directly force variations in the rate of decompression melting $\Gamma_P$, emphasising the significance of the internal-melting mechanism.
The negative flux excursions are more pronounced than the positive ones. This reflects the fact that glacial cycles are marked by gradual decreases in sea level (slow glaciation) and sharp increases (rapid collapse of ice sheets). The carbon flux variation (fig.~\ref{fig:Column_TimeSeriesSL}d) is dominantly caused by the variation in melt flux (fig.~\ref{fig:Column_TimeSeriesSL}c), rather than by the variation in carbon concentration. However, the carbon concentration plays a mitigating role, such that the carbon flux variation is about half the melt flux variation. Thus estimates of variation in melt flux from an observed variation in crustal thickness could be used to estimate variation in mid-ocean ridge carbon emissions. Figure~\ref{fig:Column_TimeSeriesSL}e,f shows that the peaks of melt and carbon flux lag the forcing by about 5~kyr. Figure~\ref{fig:Column_SLrecord_varQ}b,d shows that this estimates is sensitive to the melt transport and also influenced by 2d effects, particularly if lateral melt focussing to the ridge axis is slow.  

Carbon is a significant greenhouse gas and so any variation in magmatic carbon flux can potentially act as a feedback on glacial cycles. Our model, driven by Pleistocene fluctuations of sea level (figure~\ref{fig:Column_TimeSeriesSL}), indicates that rapid deglaciations were followed by a significant decrease in the carbon flux from mid-ocean ridges, which in turn would have reduced atmospheric carbon, a potentially significant negative climate feedback.  Because this feedback is nonlinear, the periods of forcing and response can be different. Indeed, \citet{huybers2017delayed} argued that a $\sim$120~kyr glacial cycle could arise from the 41-kyr orbital forcing when the lag of magmatic carbon emissions is 10--50~kyr. For \citet{burley2015variations}, variation in MOR emissions is driven by changes in the basal carbon flux, which is caused by variation in the depth of first melting. This produces lags of about 50--80~kyr, controlled by the melt migration rate across the whole melting column. According to \citet{huybers2017delayed}, such lags would promote longer glacial cycles than those of the Pleistocene epoch. In the present work, we account for internal melting throughout the column. This leads to a shorter lag of about 5~kyr with reference parameters (figure~\ref{fig:Column_TimeSeriesSL}f), which is much less than the melt travel time.  Figure~\ref{fig:Column_SLrecord_varQ}d shows that the lag could be even shorter if the melt velocity is faster than our preferred value, or up to about 20~kyr if the maximum melt velocity is slower (1~m/yr) and/or lateral melt focusing is slow enough. Hence, at least for the upper range of plausible lags that we predict, the mechanism proposed by \citet{huybers2017delayed} is viable. It would be interesting to revisit these feedbacks with our revised model of carbon fluxes from mid-ocean ridges. 

Our prediction of lag for mid-ocean-ridge emissions mainly depends on the segregation rate. Assuming melt ascent by diffuse porous flow, microstructural measurement of the permeability of rocks \citep[e.g.,][]{miller2014experimental} suggests speeds of the order of 1~m/yr. However, observations of U-series disequilibria are consistent with melt ascent speeds of several tens of metres per year \citep{rubin1988226ra, stracke06}. And the magmatic response in Iceland to the last deglaciation indicates rates of 50~m/yr or higher \citep{Maclennan2002, Prin2019}. Our lag predictions also depend on the rate of mantle upwelling. Current full-spreading rates range from roughly 1 to 15 cm/yr globally \citep{bown1994variation}, inducing mantle upwelling rates in the range 0.5--10~cm/yr. At our reference permeability scale, this range of mantle upwelling rates would produce lags within the range given in Fig.~\ref{fig:Column_SLrecord_varQ}. Overall, our models show that the lag is always less than about 20\% of the melt travel time. 

The theory developed here and in previous work calls out for a test by comparison with observations.  Unfortunately, time-series of lava compositions with appropriate duration and resolution are unavailable (although see \citet{Ferguson2017timeseries}). Moreover, it is unlikely that carbon dioxide would leave any observable, temporal signal after degassing. However, records of hydrothermal elemental fluxes in sediments provide a proxy for temporal variations in hydrothermal activity and ultimately in magmatic budgets over the last glacial cycles \citep{lund2011does}. The hydrothermal proxies in sediments can be accurately dated and the timing of peak hydrothermal activity can be compared to predictions from our theory. \citet{lund2016enhanced} reported time-series of Fe and Mn fluxes in the sediments of the Southern East Pacific Rise at 11$^\circ$S over the last 200 ka. They showed that the peaks in these fluxes lag the two previous maxima in the rate of sea-level decrease by about 15 kyr. \citet{middleton2016hydrothermal} found that sediments from 26$^\circ$N on the Mid-Atlantic Ridge document an increase in elemental fluxes (Fe, Cu) in hydrothermal systems concomitant with the most rapid sea-level decrease leading to the last glacial maximum. Relatively short lags between peaks in rate of sea-level change and peaks in melt flux are consistent with our models forced with the sea-level reconstruction of the Middle and Upper Pleistocene. Furthermore, consistent with the control by parameter $\mathcal{Q}$ on the lag of melt flux (figure~\ref{fig:admittance_wet}, see also figure~\ref{fig:Column_SLrecord_varQ}), the higher lags observed at the Southern East Pacific Rise might be due to the greater half-spreading rate there, compared to that of the Mid-Atlantic Ridge. 

We found differences between one- and two-dimensional models only at small forcing frequency. However, our pseudo-2d models are limited in that they do not consider potential complexities of flow focusing to the axis or lateral variations within the melting region. The latter could be caused by changes in mantle upwelling rate or melt-localisation instabilities \citep[e.g.,][]{kelemen95, keller2017volatiles, reesjones2018reaction}. To improve our understanding of the relationships between glacial cycles and mid-ocean ridge magmatism, next steps should include forcing two-dimensional models of MOR magmatism with local reconstructions of sea-level variations. They should also include comparisons with time-series of hydrothermal activity, trace elements in basaltic glass chips, and oceanic crustal thickness. 

\section*{Acknowledgements} The authors thank P.~Asimow and P.~Huybers for insightful reviews and J.F.~Rudge for comments on an early version. This research received funding from the European Research Council under the European Union's Seventh Framework Programme (FP7/2007--2013)/ERC grant agreement number 279925 and under Horizon 2020 research and innovation programme grant agreement number 772255. N.G.C.~acknowledges support from the University of Montpellier and public funding through ANR under the ``Investissements d$'$avenir'' programme with the reference ANR-16-IDEX-0006. D.R.J.~acknowledges research funding through the NERC Consortium grant NE/M000427/1, NERC Standard grant NE/I026995/1 and the Leverhulme Trust. We thank the Isaac Newton Institute for Mathematical Sciences for its hospitality during the programme Melt in the Mantle that was supported by EPSRC Grant Number EP/K032208/1. We thank the Deep Carbon Observatory of the Sloan Foundation.

\appendix
\section{Steady state for fixed sea level, parameter estimation}
\label{app:steady} 
The steady-state porosity and carbon concentration are obtained by taking the steady state of equation~\eqref{eq:full}, integrating once with respect to $z$, and applying boundary conditions~\eqref{eq:bcs1}. We find
\begin{linenomath*} \begin{subequations}
\begin{align}
    & \overline{Q} =  \Gamma^*z + \overline{c}-\mathcal{M}, \\
    & \overline{c} \left(D + \overline{Q}  \right) = D \mathcal{M},
\end{align}
\end{subequations} \end{linenomath*}
where $\overline{Q}=Q(\overline{\phi})= \mathcal{Q} \overline{\phi}^n (1-\overline{\phi})^2   + \overline{\phi}$. These algebraic equations can be readily solved; an example solution is shown in figure~\ref{fig:base}.  

The steady-state melting model can be analyzed to understand its basic behaviour. In turn, this analysis can be used to estimate the dimensionless parameters of the system in terms of (relatively) easy-to-measure quantities. 
The porosity in the melting column is controlled by the balance of melt production and melt extraction. As melt is generated, the porosity increases but the carbon concentration decreases because carbon partitions as an incompatible element into the melt. We can estimate the porosity $\phi_\mathrm{max}$ and carbon concentration $c_\mathrm{min}$ at the top of the column $z=1$ by assuming that the porosity is small and carbon is very incompatible. We define precise conditions necessary for these limits below. Under these approximations
\begin{linenomath*} \begin{subequations}
\begin{align}
    &  \mathcal{Q}  \phi^n_\mathrm{max} \sim  \Gamma^* - \mathcal{M}, \quad \Rightarrow \quad \phi_\mathrm{max} \sim \left(\frac{\Gamma^* - \mathcal{M} }{\mathcal{Q}}\right)^{1/n}, \label{eq:phimaxG}\\
    &  c_\mathrm{min} \sim \frac{D \mathcal{M}}{\Gamma^* - \mathcal{M}}.
\end{align}
\end{subequations} \end{linenomath*}
The corresponding concentration of the solid and liquid phases, respectively are 
\refstepcounter{equation}
\[
    c_{s,\mathrm{min}} \sim {D }/(\Gamma^* - \mathcal{M}), \qquad  
    c_{l,\mathrm{min}} \sim {1 }/(\Gamma^* - \mathcal{M}) .
    \eqno{(\theequation{\mathit{a},\mathit{b}})} 
\]
We then determine appropriate conditions under which the approximations hold by estimating \textit{a posteriori} the magnitude of each of the neglected terms. In particular, we require
\refstepcounter{equation} \label{eq:constraints}
\[
    \mathcal{Q} \gg \left(\Gamma^* - \mathcal{M}\right)^{1-n}, \qquad  
    D \ll \left\{\Gamma^* - \mathcal{M},\frac{(\Gamma^* - \mathcal{M})^2}{\mathcal{M}} \right\} .
    \eqno{(\theequation{\mathit{a},\mathit{b}})} 
\]
The first expression ensures that the term $+\overline{\phi}$ can be neglected in the flux $\overline{Q}$. Then $\overline{\phi}\ll \Gamma^*-\mathcal{M}<1$. The second expression ensures that $\overline{c}\ll \Gamma^*-\mathcal{M}$. Physically, these conditions mean that the melt extraction is fast enough to maintain a small porosity and the compatibility is low enough that carbon does not suppress the solidus much near the top of the column.

We now relate our governing parameters to other quantities of interest, which may be easier to measure or estimate in practice. First, in the limit of small porosity, the maximum flux is equivalent to the maximum degree of melting $F_\mathrm{max}$ \citep{Ribe1985}. In particular,
\begin{equation}
    F_\mathrm{max} \sim \Gamma^* - \mathcal{M}.
\end{equation}
So, using equation~\eqref{eq:phimaxG}, the maximum porosity satisfies 
\begin{equation}
    \phi_\mathrm{max} \sim \left(\frac{F_\mathrm{max} }{\mathcal{Q}}\right)^{1/n} = \left(\frac{F_\mathrm{max} \mu W_0 }{k \Delta \rho g}\right)^{1/n},
\end{equation}
Recall that $\mathcal{M}$ is related to $\Gamma^*$ through equation~\eqref{eq:M(Gamma)}, so
\refstepcounter{equation}
\[
    \Gamma^* \sim F_\mathrm{max} \frac{H}{H_\mathrm{dry}}, \qquad  
    \mathcal{M} \sim  F_\mathrm{max} \left(\frac{H}{H_\mathrm{dry}} -1 \right).
    \eqno{(\theequation{\mathit{a},\mathit{b}})} 
\]
For example, we take the following broadly accepted values from the literature: $F_\mathrm{max} \approx 0.2$, the carbonated solidus is at $H\approx130$~km, and the dry solidus is at $H_\mathrm{dry}\approx65$~km. Then $\Gamma^*=0.4$ and $\mathcal{M}=0.2$ \citep[e.g.,][]{klein1987global, Forsyth1998MELT}.

Second, we estimate the liquid velocity at the top of the column 
\begin{equation} \label{eq:w_ratio_app}
    \frac{w_0}{W_0} \sim  \frac{F_\mathrm{max}}{\phi_\mathrm{max}} \sim F_\mathrm{max}^{\quad \tfrac{n-1}{n}}\mathcal{Q}^{\tfrac{1}{n}}.
\end{equation}
Given an estimate of the melt velocity (say from observations of uranium-series disequilibrium \citep{rubin1988226ra, stracke06} or the Iceland post-glacial melt pulse \citep{Maclennan2002,Swindles2017,Prin2019}), we can estimate 
\begin{equation}
  \mathcal{Q} =  (w_0/W_0)^{ n}   F_\mathrm{max}^{1-n}.
\end{equation}
For example, if the melt velocity is 140 times faster than the mantle upwelling rate, $w_0/W_0=140$. If also $n=2$, which is appropriate given that the porosity is small \citep{rudge18}, then $\mathcal{Q} = 10^5$. As expected, $\mathcal{Q}$ is relatively large since melt extraction is fast and porosities remain small. Also, $D\sim 10^{-4}$ for carbon \citep{rosenthal2015experimental} so all the constraints required in equation~\eqref{eq:constraints} are satisfied.

Despite the consensus on the physical quantities above (similar parameters are used in previous studies \citep{crowley2015glacial,burley2015variations}) there is uncertainty arising both from the indirect nature of the constraints and from geographical variation.  This is particularly true of $\mathcal{Q}$, which depends on permeability, spreading rate and melt viscosity. We explore model sensitivity to the choice of $\mathcal{Q}$ in figures~\ref{fig:admittance_wet} and \ref{fig:Column_SLrecord_varQ} and the Supplementary Information (section S1).

\section{Effect of fluctations in sea level} \label{sec:perturbed}
The equations governing the time-dependent fluctuations are obtained by linearizing equation~\eqref{eq:full} about the steady part $(\overline{\phi},\,\overline{c})$ of the solution. In particular, we neglect terms that contain $\delta_0^2$ or higher powers. We collect terms proportional to $\delta_0$ and find
\begin{linenomath*} \begin{subequations} \label{eq:pert}
\begin{align}
    &\frac{d \hat{Q}}{d z} =  
    i\omega\left(-\Gamma^* -\hat{\phi} + \hat{c} \right)
    +\frac{d \hat{c}}{d z}, \label{eq:pert-a} \\
    & \left( D + \overline{Q} + \overline{c} \right) \frac{d \hat{c}}{d z}  
    = i\omega\overline{c}\Gamma^*
    -\hat{c} \left[i\omega\left(D+\overline{\phi}+\overline{c} \right) +\frac{d \overline{Q}}{d z}   \right]
    -\hat{Q}\frac{d \overline{c}}{d z}  . \label{eq:pert-b}
\end{align}
\end{subequations} \end{linenomath*}
Note that $\hat{Q} = \hat{\phi}\,  \tfrac{d\overline{Q}}{d\overline{\phi}} $. Appropriate boundary conditions are found by linearizing equation~\eqref{eq:bcs1} to obtain
\refstepcounter{equation}
\[
    \hat{Q}(z=0) =- \left. \frac{d \overline{Q}}{dz} \right|_{z=0},   \qquad  
    \hat{c}(z=0) =- \left. \frac{d \overline{c}}{dz} \right|_{z=0}  .
    \eqno{(\theequation{\mathit{a},\mathit{b}})} \label{eq:pert-bcs1}
\]
Hence equation~\eqref{eq:pert} is a pair of coupled ordinary differential equations that can be solved for $\hat{Q}$ (and hence $\hat{\phi}$) and $\hat{c}$.

We also develop a `basal-flux' model akin to that of \citet{burley2015variations}. In this model, we neglect the contributions from internal melting. In particular, we neglect the term $-i\omega\Gamma^*$ in equation~\eqref{eq:pert-a}  and $i\omega\overline{c}\Gamma^*$ in equation~\eqref{eq:pert-b}. Results using this model are presented in figure~\ref{fig:DryBasal} and discussed in section~\ref{sec:Discussion}.

\section{Approximate solutions valid in the limit of large forcing frequency} \label{sec:large_omega}

When the period of sea-level fluctuation is short, the frequency $\omega$ is large. In this limit, we derive approximate solutions of the equations given in Appendix~\ref{sec:perturbed}. These approximate solutions are useful in that they allow us to identify the physical mechanisms of importance in this regime (as described in the main text), as well as simple estimates of quantities of interest, such as the melt and carbon flux. 

The key methodological idea is that the imaginary part of the fluctuating quantities is very much smaller than the real part, by a factor of $\omega^{-1}\ll 1$. Mathematically, this can be seen by inspection of equation~\eqref{eq:pert} (we discuss the physical meaning in section~\ref{sec:results_period}). When $\omega$ is very large, the collective group of terms multiplied by it must be very small. This then shows that $\hat{\phi}$ and $\hat{c}$ must be (approximately) equal to some real function of $\Gamma^*$, $D$, $\overline{\phi}$ and $\overline{c}$, all of which are real. So the fluctuations are approximately equal to their real parts. Then it can be seen that the imaginary parts are a factor $\omega^{-1}\ll 1$ smaller. With this in mind, we write:
\refstepcounter{equation}
\[
    \hat{Q} = \hat{Q}_r + i \omega^{-1} \hat{Q}_i,   \qquad 
    \hat{\phi} = \hat{\phi}_r + i \omega^{-1} \hat{\phi}_i,   \qquad 
    \hat{c} = \hat{c}_r + i\omega^{-1} \hat{c}_i  .
    \eqno{(\theequation{\mathit{a},\mathit{b},\mathit{c}})} \label{eq:pert-decomposition}
\]
In this expansion, a subscript $r$ denotes the real part of a quantity and a subscript $i$ denotes the imaginary part scaled with the frequency. 
Thus, for example, both $\hat{Q}_r$ and $\hat{Q}_i$ are real quantities.
We substitute this decomposition into equation~\eqref{eq:pert}, take the real and imaginary parts, and collect powers of $\omega$.

The leading order balances, in which we collect terms proportional to $\omega$, gives
\begin{linenomath*} \begin{subequations} \label{eq:pertD0}
\begin{align}
    &0 =  -\Gamma^* -\hat{\phi}_r + \hat{c}_r 
    , \label{eq:pertD0-a} \\
    & 0 = \overline{c}\Gamma^*
    -\hat{c}_r \left(D+\overline{\phi}+\overline{c} \right) . \label{eq:pertD0-b}
\end{align}
\end{subequations} \end{linenomath*}
These expressions can be straightforwardly rearranged to give explicit solutions for the real part of all the fluctuations in terms of the mean state. In particular
\begin{linenomath*} \begin{subequations} \label{eq:pertD0R}
\begin{align}
    &\hat{\phi}_r = -\frac{D+\overline{\phi}}{D+\overline{\phi}+\overline{c}} \Gamma^* 
    , \label{eq:pertD0R-a} \\
    & \hat{c}_r = \frac{\overline{c}}{D+\overline{\phi}+\overline{c}}  \Gamma^*
     . \label{eq:pertD0R-b}
\end{align}
\end{subequations} \end{linenomath*}
Note that $\hat{Q}_r = \hat{\phi}_r\, \tfrac{d\overline{Q}}{d\overline{\phi}} $.  

Then, the next order balances, in which we collect terms proportional to $\omega^0$, gives
\begin{linenomath*} \begin{subequations} \label{eq:pertD1}
\begin{align}
    &\frac{d \hat{Q}_r}{d z} =  
    \left( \hat{\phi}_i - \hat{c}_i \right)
    +\frac{d \hat{c}_r}{d z}, \label{eq:pertD1-a} \\
    & \left( D + \overline{Q} + \overline{c} \right) \frac{d \hat{c}_r}{d z}  
    =     \hat{c}_i\left(D+\overline{\phi}+\overline{c} \right)
    -\hat{c}_r \frac{d \overline{Q}}{d z}  -\hat{Q}_r\frac{d \overline{c}}{d z}  . \label{eq:pertD1-b}
\end{align}
\end{subequations} \end{linenomath*}
As before, we can rearrange these expressions to give explicit expressions for $\hat{\phi}_i$ and $\hat{c}_i$, since the terms involving real parts of the fluctuations are known from equation~\eqref{eq:pertD0R}. We can also obtain the imaginary part of the melt flux fluctuation by noting that $\hat{Q}_i = \hat{\phi}_i\,  \tfrac{d\overline{Q}}{d\overline{\phi}} $.

Our approximate expressions for the real part of all the fluctuating quantities allow us to estimate the admittance (in terms of melt and carbon fluxes), since the full fluctuation is dominated by the real part. 
The admittance of melt flux (in units of per 100~m peak-to-trough sea-level fluctuation) is 
\begin{equation}
    A_Q = \frac{\delta_0 \Gamma^*}{\overline{Q}}  \frac{D+\overline{\phi}}{D+\overline{\phi}+\overline{c}} \frac{d\overline{Q}}{d\overline{\phi}},
\end{equation}
where all quantities are evaluated at the top of the melting column (and we convert to a percentage when plotting).
In~\ref{app:steady}, we discussed how the steady-state variables can be simplified at the top of the melting column. If we make the approximation that \mbox{$D \ll \overline{\phi} \ll 1$} and \mbox{$\overline{c} \ll \overline{\phi} $}, then
\begin{equation} \label{eq:admittance_melt_approx}
    A_Q \sim n \frac{\Delta S}{H_\mathrm{dry}}  \frac{\rho_w}{\rho} F_\mathrm{max}^\frac{n-1}{n} \mathcal{Q}^\frac{1}{n} \sim n \frac{\Delta S}{H_\mathrm{dry}}  \frac{\rho_w}{\rho} \frac{w_0}{W_0},
\end{equation}
where $\Delta S$ is the dimensional magnitude of the sea-level fluctuation and $w_0$ is the dimensional melt velocity at the top of the column. Thus the admittance is proportional to the pressure fluctuation induced by sea level relative to static pressure over the dry melting region (this quantity is typically very small). However, it is also multiplied by $w_0/W_0$, the ratio of melt velocity at the top of the column to the mantle upwelling velocity, which is typically fairly large. Thus the admittance can be significant. The admittance of porosity is smaller than that of melt flux by a factor of $\tfrac{1}{n}$, since we lose the factor coming from $\tfrac{d\overline{Q}}{d\overline{\phi}}$. 

The admittance of the carbon flux $A_{Q_c}$ can be estimated
\begin{equation} 
A_{Q_c} = {\delta_0 \Gamma^*}\left( \frac{D+\overline{\phi}}{D+\overline{\phi}+\overline{c}}\frac{d\overline{Q}}{d\overline{\phi}}  \frac{1 }{\overline{Q}} - \frac{1}{D+\overline{\phi}+\overline{c}} \right)
\end{equation}
where the first term comes from the fluctuation in melt flux and the seccond term comes from the fluctuation in carbon concentration.
By making the same approximations to the steady state used to derive~\eqref{eq:admittance_melt_approx}, we find that the admittance of carbon flux is smaller than that of melt flux by a factor $\tfrac{n-1}{n}$. In particular, 
\begin{equation} \label{eq:admittance_carbon_approx}
    A_{Q_c} \sim (n-1) \frac{\Delta S}{H_\mathrm{dry}}  \frac{\rho_w}{\rho} F_\mathrm{max}^\frac{n-1}{n} \mathcal{Q}^\frac{1}{n} \sim (n-1) \frac{\Delta S}{H_\mathrm{dry}}  \frac{\rho_w}{\rho} \frac{w_0}{W_0}.
\end{equation}
The admittance of carbon flux is less than that of the melt flux because the variation in carbon concentration partially compensates the variation in melt flux. While these predictions only apply to the short period regime, they are consistent with the wider pattern observed in figure~\ref{fig:Column_TimeSeriesSL} for the sea-level record over the past 800~kyr. Admittance of the porosity is half that of the melt flux which is about double that of the carbon flux, consistent with the above theory when $n=2$.

Moreover, by calculating the imaginary part of the fluctuations, we can also estimate the phase of the fluctuation. Finally, by comparing the magnitude of the real and imaginary parts, we can estimate the critical frequency below which this large $\omega$ regime no longer applies. Physically, this allows us to estimate when the fluxes are proportional to sea level and when they are proportional to the rate of change of sea level. Both of these depend only on the ratio of the scaled imaginary part of the fluctuation to the real part. Approximate formulae for these are
\begin{linenomath*} \begin{subequations} \label{eq:pert_ratio}
\begin{align}
    &\frac{\hat{\phi}_i}{\hat{\phi}_r} =  (n-1)\Gamma^* \left(\frac{\mathcal{Q}}{F_\mathrm{max}} \right)^{1/n}     , \label{eq:pert_ratio-a} \\
    & \frac{\hat{c}_i}{\hat{c}_r} =  \left(n-\tfrac{1}{n}\right)\Gamma^* \left(\frac{\mathcal{Q}}{F_\mathrm{max}} \right)^{1/n}  ,  \label{eq:pert_ratio-b} \\
    & \frac{\hat{Q}_{ci}}{\hat{Q}_{cr}} =  \frac{n^2-2n + \tfrac{1}{n}}{n-1}\Gamma^* \left(\frac{\mathcal{Q}}{F_\mathrm{max}} \right)^{1/n}  . \label{eq:pert_ratio-c}
\end{align}
\end{subequations} \end{linenomath*}
Note that equation~\eqref{eq:pert_ratio-a} also applies to melt flux, since $\hat{Q}_i/\hat{Q}_r=\hat{\phi}_i/\hat{\phi}_r$ . In deriving equation~\eqref{eq:pert_ratio-b}, we assumed that \mbox{$D\ll \overline{\phi} \Gamma^{*2} / \mathcal{M} F_\mathrm{max}$ }. Finally, we can use equation~\eqref{eq:pert_ratio} to infer the critical period above which the phase shift is significant. We find a dimensional critical period for the melt flux and carbon fluxes, respectively, 
\begin{linenomath*} \begin{subequations} \label{eq:critical_tp_app}
\begin{align}
& t_p^* \sim  C  \frac{\hat{\phi}_r}{\hat{\phi}_i}    \frac{H}{W_0 } \sim  C  \frac{1}{n-1} \frac{H_\mathrm{dry}}{w_0}, \label{eq:critical_tp_app-Q} \\
& t_p^* \sim  C  \frac{\hat{Q}_{cr}}{\hat{Q}_{ci}}    \frac{H}{W_0 } \sim  C  \frac{n-1}{n^2-2n + \tfrac{1}{n}} \frac{H_\mathrm{dry}}{w_0}, \label{eq:critical_tp_app_Qc}
\end{align}
\end{subequations} \end{linenomath*}where $C\approx 1$ is a dimensionless prefactor (which can be chosen to match a specific phase shift). For the parameters given in table~\ref{tab:parameters}, the critical period for the melt flux is about 23~kyr, consistent with figure~\ref{fig:admittance_wet}(b). Note that ${H_\mathrm{dry}}/{w_0}$ is a measure of the transit time of melt across the dry melting region. The transit time based on the true $z$-dependent velocity is (approximately, when $\overline{\phi} \ll 1$) a factor of $n$ larger than the time based on the maximum melt velocity  $w_0$. If the forcing period is much less than the melt transport time, there is little melt segregation over the forcing cycle and instead the porosity fluctuation and carbon concentration respond near instantaneously to the change in sea level.

\section{Lateral melt focusing in a pseudo-two-dimensional melting region}
\label{sec:focusing_model}
Following \cite{langmuir1992petrological}, we consider a triangular melting region with a base at $z=0$ and apex at the ridge axis, $z=1$. As in the main text, all lengths are non-dimensionalized by $H$ and fluxes by $W_0$. The melting region is capped by a decompaction channel with dip $\alpha$ \citep{sparks1991melt}, sketched in figure~\ref{fig:sketch}. At each distance $x$ from the axis, a melting column spans $0 \le z \le z_c(x)$, where $z_c(x) \equiv 1 - x\tan\alpha$ is the depth to the decompaction channel as a function of distance $x$ off-axis. All melting columns (except the one at $x=0$) empty into the decompaction channel, which focuses magma laterally up-dip to the ridge axis. Only magma that enters the decompaction channel within some maximum focusing distance $x_f$ actually arrives at the ridge axis \citep[e.g.,][]{katz2008magma, hebert10}.  Following \cite{burley2015variations}, we choose $x_f$ to give a mean crustal thickness of 7~km at the ridge axis.

To keep things simple (and consistent with the triangular geometry), we assume a uniform upwelling rate at the bottom of the melting region for all $x$. Hence all columns are identical except for their height $z_c$. The rate $Q_\text{mor}$ at which magma arrives at the ridge axis is
\begin{equation}
    Q_\text{mor}(t) = 2\int_0^{x_f} Q[1-h(x),t-{\tau}(x)]\,d x,
\end{equation}
where ${\tau}(x)$ is the travel time required for melts that enter the decompaction channel at a distance $x$ to arrive at the ridge axis. The factor of two comes from the contributions from both sides of the axis, assuming mirror symmetry about $x=0$. Note that $Q_\text{mor}$ is a flux multiplied by a length, so the dimensional equivalent needs to be multiplied by a factor of $W_0 H$. We use the same method for the carbon flux.

Then, writing $ dx = -d{z}/\tan\alpha$, we express the integral as 
\begin{equation}
    Q_\text{mor}(t) = \frac{2}{\tan\alpha}\int_{z_c(x_f)}^1 Q[z,t-{\tau}(z)]\,dz.
\end{equation}
Note that $\tau(z)=\tau(z_c(x))$, where $z_c(x)$ is the top of the melting column at position $x$.  It is understood that regardless of the argument ($x$ or $z_c)$, $\tau$ represents the transit time of magma laterally along the sloping decompaction channel.

The choice of ${\tau}$ closes the model. We consider two cases. First, we take $\tau=0$ to represent instantaneous focussing as in \cite{burley2015variations}. Second, we follow \cite{crowley2015glacial} in using the steady-state column transport time to define the focusing transport time. In particular, we assume that the focusing time of melt that enters the decompaction channel is the same as the time that would be required for melt to continue up the melting column to the top. Thus
\begin{equation}
    \label{eq:focusing-time}
    {\tau}(x) = \int_{z_c(x)}^1 \frac{dz}{ \overline{w}(z)}\,.
\end{equation}

For either case, we use the decomposition \eqref{eq:decomposition-defintion} to separate the MOR magma delivery rate into steady and fluctuating parts as
\begin{equation}
    \label{eq:focusing-model}
    Q_\text{mor}(t) = \frac{2}{\tan\alpha}\left\{\int_{z_f}^1 \overline{Q}(z)\,dz + \delta_0 \int_{z_f}^1\hat{Q}(z)\exp{\left[i\omega(t-{\tau}(z))\right]}\,dz \right\}.
\end{equation}
where $z_f \equiv z_c(x_f) = 1 - x_f\tan\alpha$.

Note that the dimensional mean crustal thickness is given by
\begin{equation}
    \overline{H}_\text{crust} = \frac{H W_0}{U_0\tan\alpha} \frac{\rho_l}{\rho_c} \int_{z_f}^1 \overline{Q}(z)\,dz 
\end{equation}
where $\rho_l=2800$ kg~m$^{-3}$ and $\rho_c=2900$ kg~m$^{-3}$ are magma and crustal density, respectively.

The admittance and lag of the pseudo-2d models is computed identically to that of the column model, except in using $Q_\text{mor}$ from equation~\eqref{eq:focusing-model}. Carbon fluxes are treated in the same way.

\section*{References}
\bibliographystyle{apsrev4-2.bst}
\bibliography{bib_MORs}

\clearpage
\newpage

\noindent {\centering \textbf{{\large{SUPPLEMENTARY MATERIAL}} } }
\setcounter{section}{0}
\setcounter{figure}{0}
\setcounter{equation}{0}

\renewcommand\thefigure{S\arabic{figure}}  
\renewcommand\thesection{S\arabic{section}}  
\renewcommand\theequation{S\arabic{equation}}  

The Supplementary Material contains :
\begin{itemize}
    \item[S1.] Models showing the dependency of the fluctuating state to the fluid dynamical parameters of the system. We provide the results for models with: 
    \begin{itemize}
        \item Melt flux parameter $\mathcal{Q}=0.25\times10^{5}$. This value corresponds to that in models for which we compute the admittance in the main text (dashed lines in figure 5, panels a,b).
        \item Melt flux parameter $\mathcal{Q}=4\times10^{5}$.
        \item Exponent in permeability--porosity relationship $n=3$.
        \item Maximum degree of melting $F_{\text{max}}=0.25$.
        \item Partition coefficient for carbon $D_c=5\times10^{-4}$.
    \end{itemize}  
    \item[S2.] Plots summarising the admittance and lag of fluxes as a function of forcing period. These plots differ from those in the main text in that they make a comparison between numerical results and asymptotic theory for small- and large-period forcing. The theory is described in the main text and its Appendix~C.
\end{itemize}

\section{Effect of changes in the fluid mechanical properties of the system}
Below, we describe the behaviour of several models where we have varied the fluid mechanical properties of the system relative to the reference model described in the main text. The fluctuations are forced with a dimensional period of $t_p=100$~kyrs.\\

\begin{figure}
    \centering 
    \includegraphics[width=0.9\linewidth]{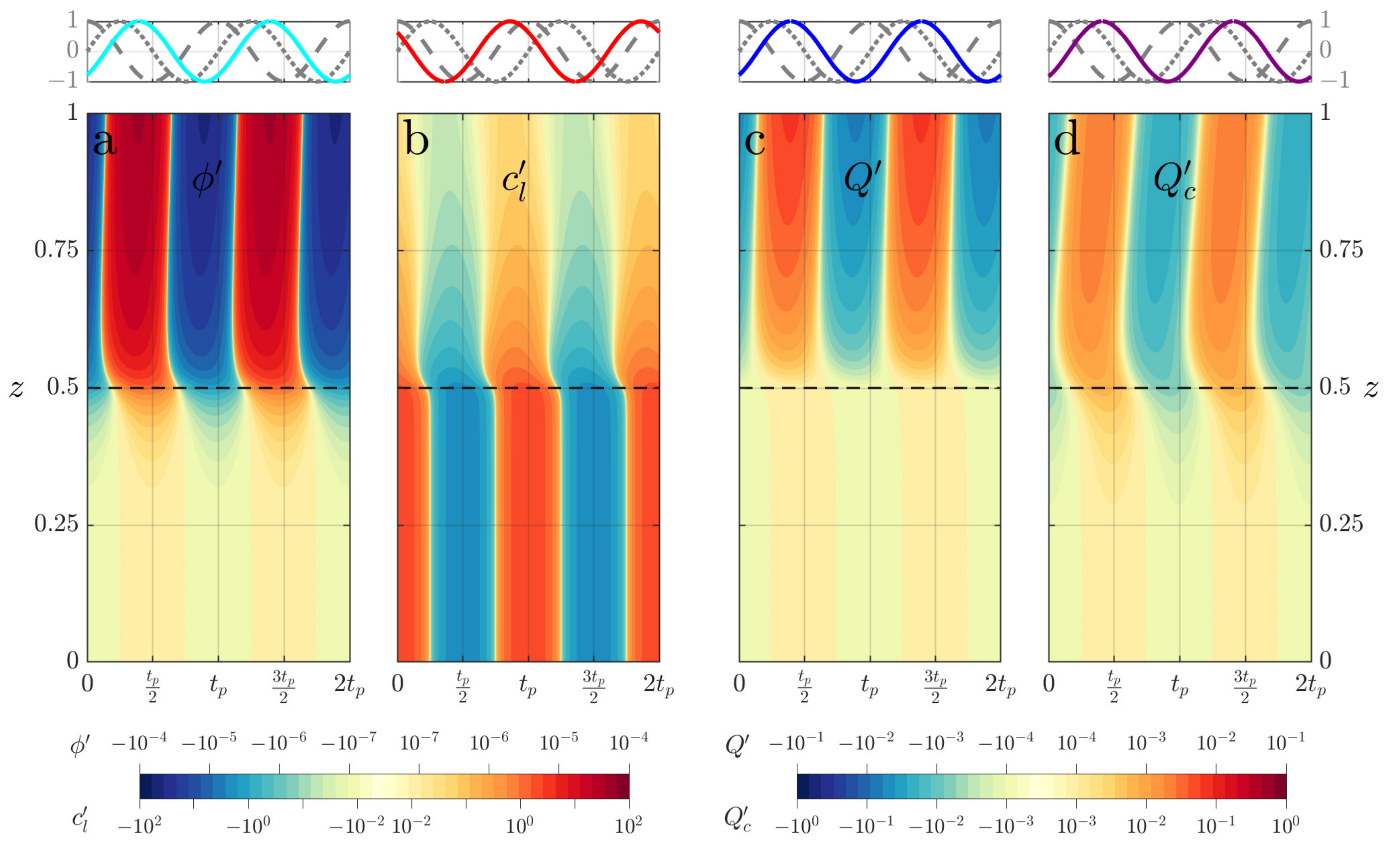}
    \caption{Time-dependent fluctuations in a model with $\mathcal{Q}=0.25\times10^{5}$. For other legend details see figure~3 in the main text.}
    \label{fig:PS_Q0.25e5}
\end{figure}

Figure \ref{fig:PS_Q0.25e5} shows the fluctuating state of a model where $\mathcal{Q}=0.25\times10^{5}$, e.g., a case with a lower permeability and/or a higher mantle upwelling velocity compared to the reference model that was described in the main text. In figure~\ref{fig:PS_Q0.25e5}, the steady-state porosity is higher because there is less melt segregation; the steady-state carbon concentration and fluxes remain identical to that in the reference model. The fluctuations in porosity and thus in melt flux are higher than in the reference model because of the lower segregation, and the fluctuations in carbon concentration are lower due to dilution. As a consequence, the fluctuating carbon flux is almost solely driven by fluctuations in melt flux. Also, as expected, the forcing period $t_p=100$~kyr used for this model is close to the critical period since the latter increases with decreasing melt velocity.\\

\begin{figure}
    \centering 
    \includegraphics[width=0.9\linewidth]{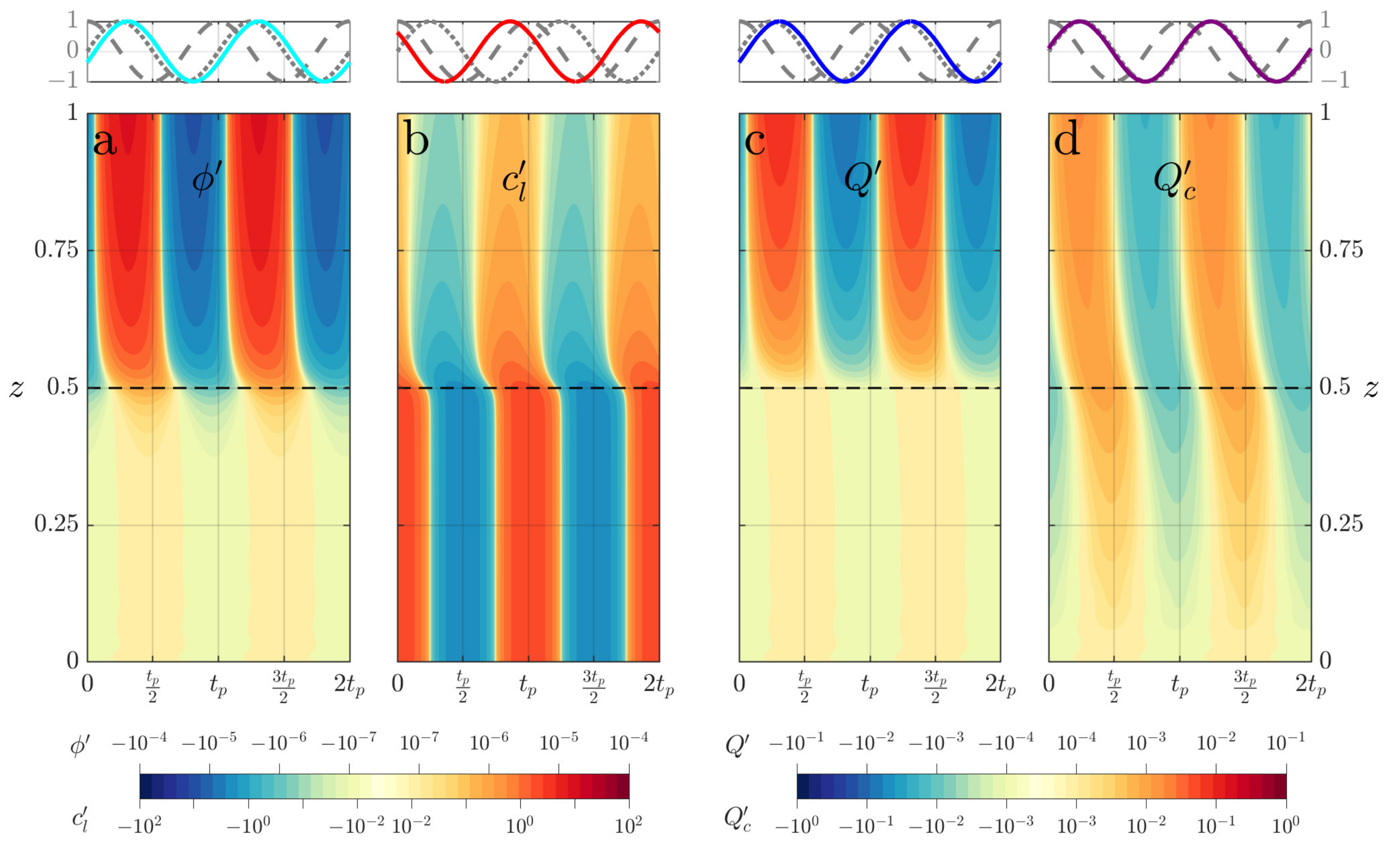}
    \caption{Time-dependent fluctuations in a model with $\mathcal{Q}=4\times10^{5}$. For other legend details see figure~3 in the main text.}
    \label{fig:PS_Q4e5}
\end{figure}

Figure \ref{fig:PS_Q4e5} shows the fluctuating state of a model where $\mathcal{Q}=4\times10^{5}$, e.g., a case with a higher permeability and/or a lower mantle upwelling velocity compared to the reference model. The trends are the opposite to that described above for the model with $\mathcal{Q}=0.25\times10^{5}$. The higher melt segregation decreases fluctuations in porosity and melt fluxes, and induces less dilution and thus higher fluctuations in carbon concentration. The effect of fluctuating carbon concentration on carbon flux is higher than in the reference model. This could also be understood as the critical period being smaller when $\mathcal{Q}=4\times10^{5}$.\\

\begin{figure}
    \centering 
    \includegraphics[width=0.9\linewidth]{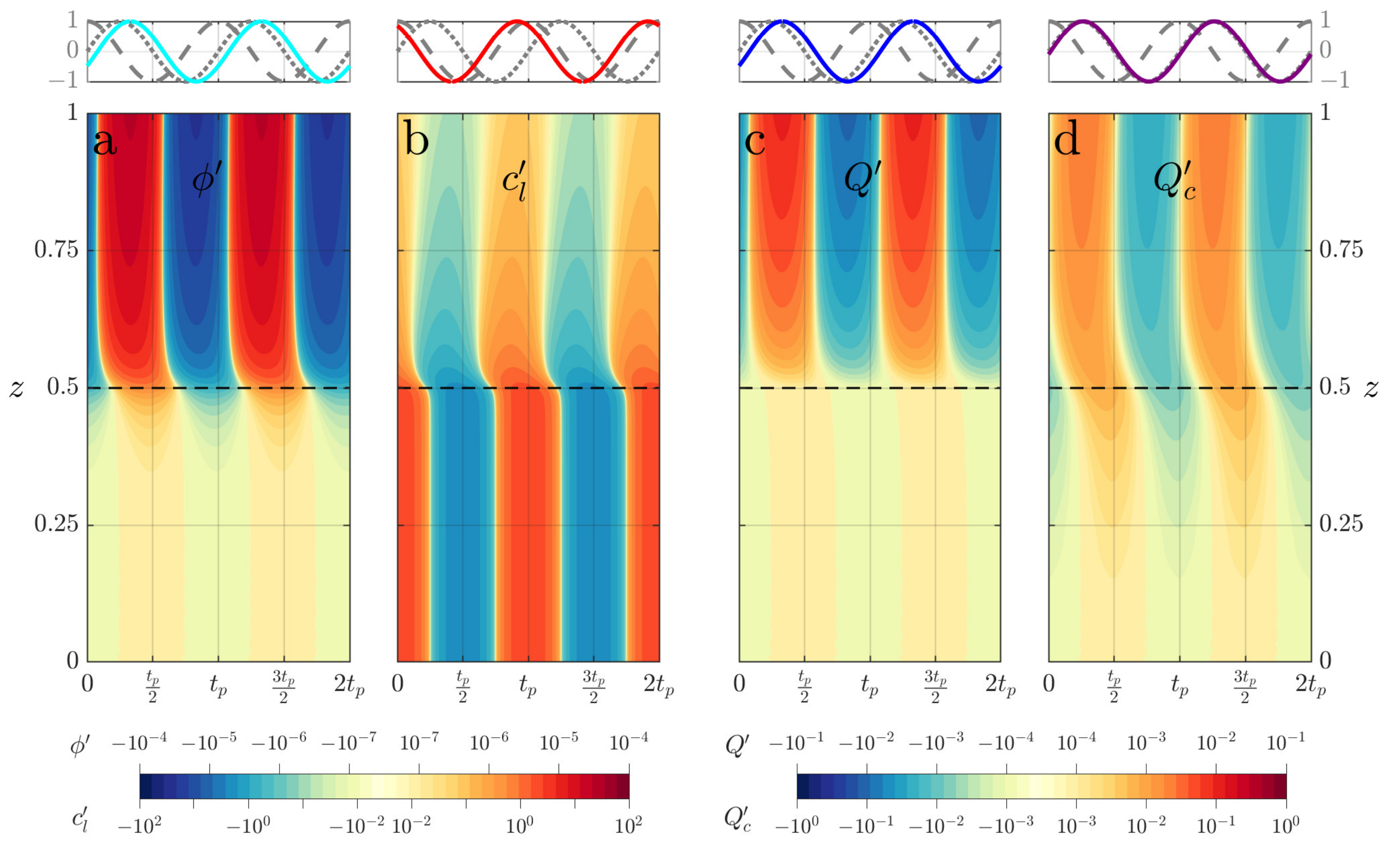}
    \caption{Time-dependent fluctuations in a model with $F_{\text{max}}=0.25$. For other legend details see figure~3 in the main text}
    \label{fig:PS_F025}
\end{figure}

Figure \ref{fig:PS_F025} shows the fluctuating state of a model where $F_{\text{max}}=0.25$. A higher degree of melting slightly increases the steady-state porosity and melt flux, and slightly decreases steady-liquid concentration by dilution. Yet, because the differences in the steady-state variables relative to that in the reference model are modest, the consequences of increasing $F_{\text{max}}$ by $25\%$ (i.e., from 0.2 to 0.25) on the fluctuations are small. \\

\begin{figure}
    \centering 
    \includegraphics[width=0.9\linewidth]{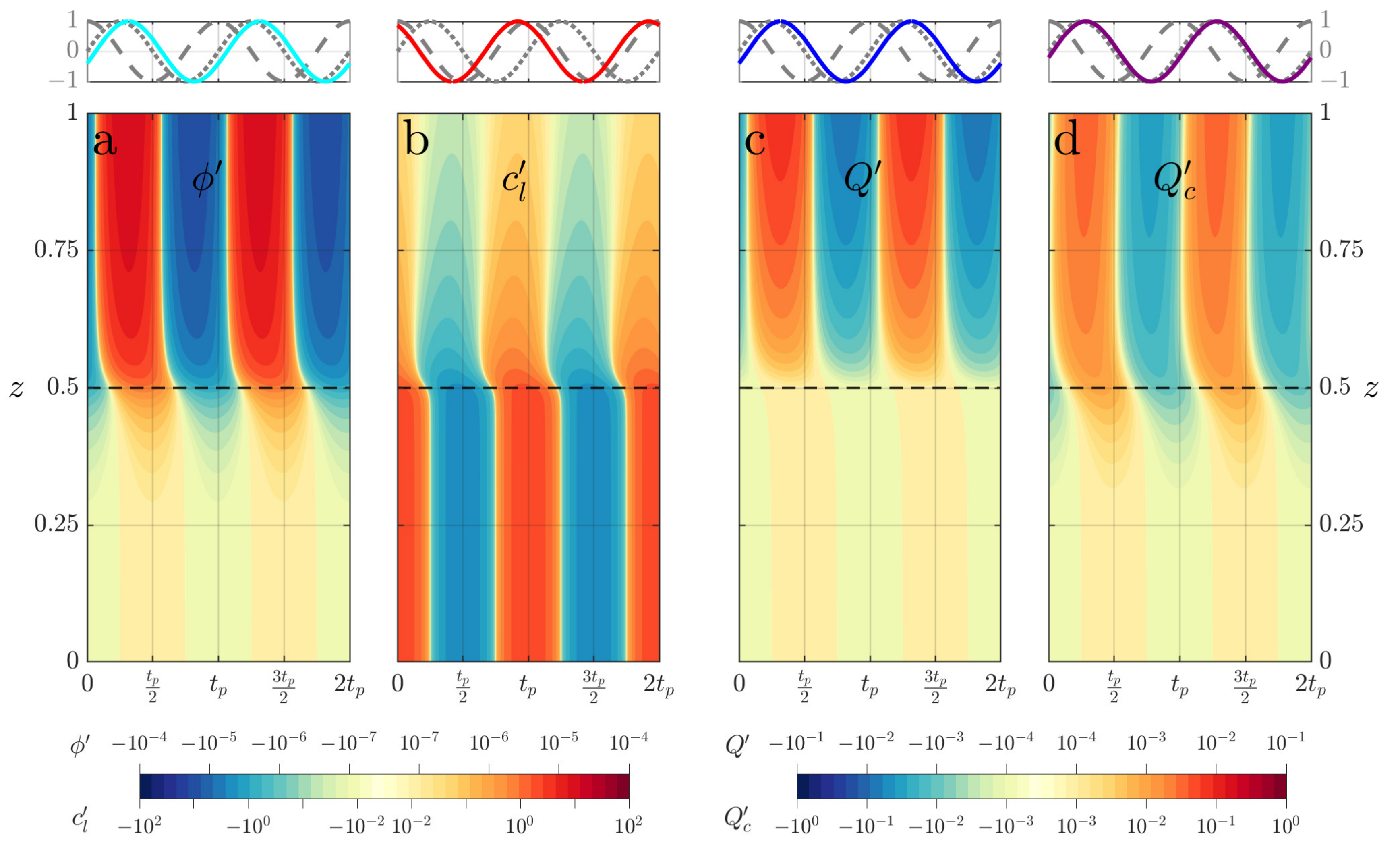}
    \caption{Time-dependent fluctuations in a model with $n=3$ and $\mathcal{Q}=7.0\times10^{7}$. For other legend details see figure~3 in the main text}
    \label{fig:PS_n3}
\end{figure}

Figure \ref{fig:PS_n3} shows the fluctuating state of a model with $n=3$. The value of $\mathcal{Q}$ is chosen to keep the melt-to-solid velocity ratio identical to that in the reference model. Increasing the exponent in the permeability--porosity relationship increases the steady-state porosity in most of the column relative to that in the reference model, while the steady concentration and fluxes remain unchanged. The fluctuations in porosity and carbon concentration are lower than in the reference model, and their phase is similar. The fluctuations in melt flux are similar and those in carbon flux, mainly driven by fluctuating melt flux, are larger in the case with $n=3$. At the top of the column, the amplitude of porosity relative to its mean steady-state value is lower with $n=3$. This ratio is similar and higher for melt and carbon flux, respectively. These results are consistent with the analysis developped in Appendix~C. Since the critical period is higher for $n=2$ ($t_p^*\simeq20$~kyr) than for $n=3$ ($t_p^*\simeq10$~kyr), the lower ratio of fluctuations in porosity to the steady-state is indeed expected in the case $n=3$. \\

\begin{figure}
    \centering 
    \includegraphics[width=0.9\linewidth]{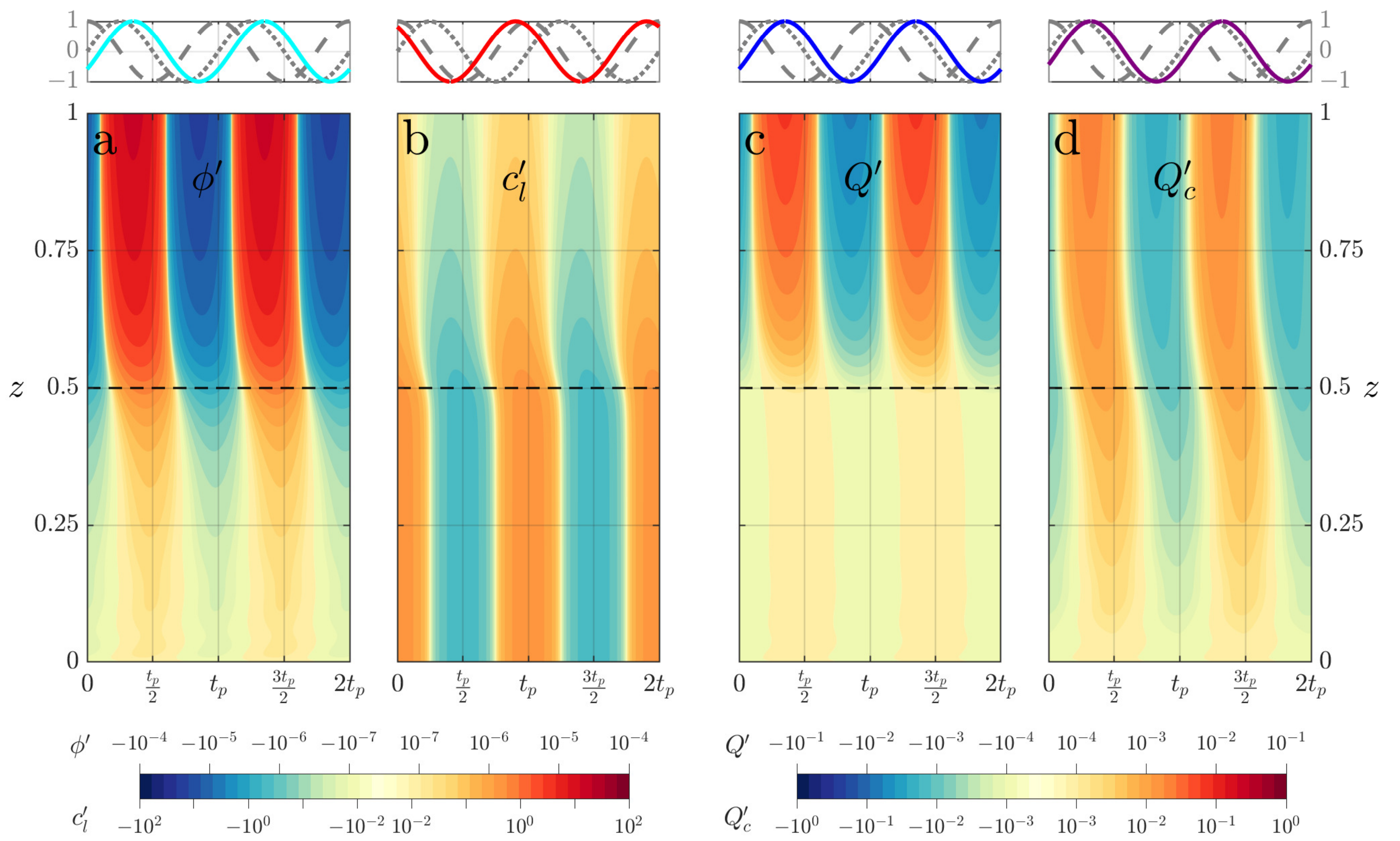}
    \caption{Time-dependent fluctuations in a model with $D_c=5\times10^{-4}$. For other legend details see figure~3 in the main text}
    \label{fig:PS_D5e-4}
\end{figure}

Figure \ref{fig:PS_D5e-4} displays the fluctuating state of a model with $D_c=5\times10^{-4}$. The higher partition coefficient of carbon induces a slower decrease of steady-carbon concentration in the solid with height, relative to that in the reference model. This promotes more melt and more melt segregation in the wet regime in the former case; hence the transitional regime starts deeper. As a consequence, in the transitional regime the fluctuating porosity is higher and fluctuating liquid concentration is lower. In the dry regime, the magnitude of fluctuations in porosity and carbon concentration are lower than in the reference model. Similarly, the fluctuating melt and carbon fluxes are slightly lower in the case of a less incompatible carbon, with differences at the top of the column of about $20-30\%$. Yet the ratio between the amplitude of fluctuations and the steady-state of fluxes does not change with $D_c$, as predicted by the approximation for an incompatible element given in Appendix~C of the main text. 

\section{Test of predictions of simplified model valid when the forcing period is very short or very long}
\begin{figure}
    \centering 
    \includegraphics[width=0.8\linewidth]{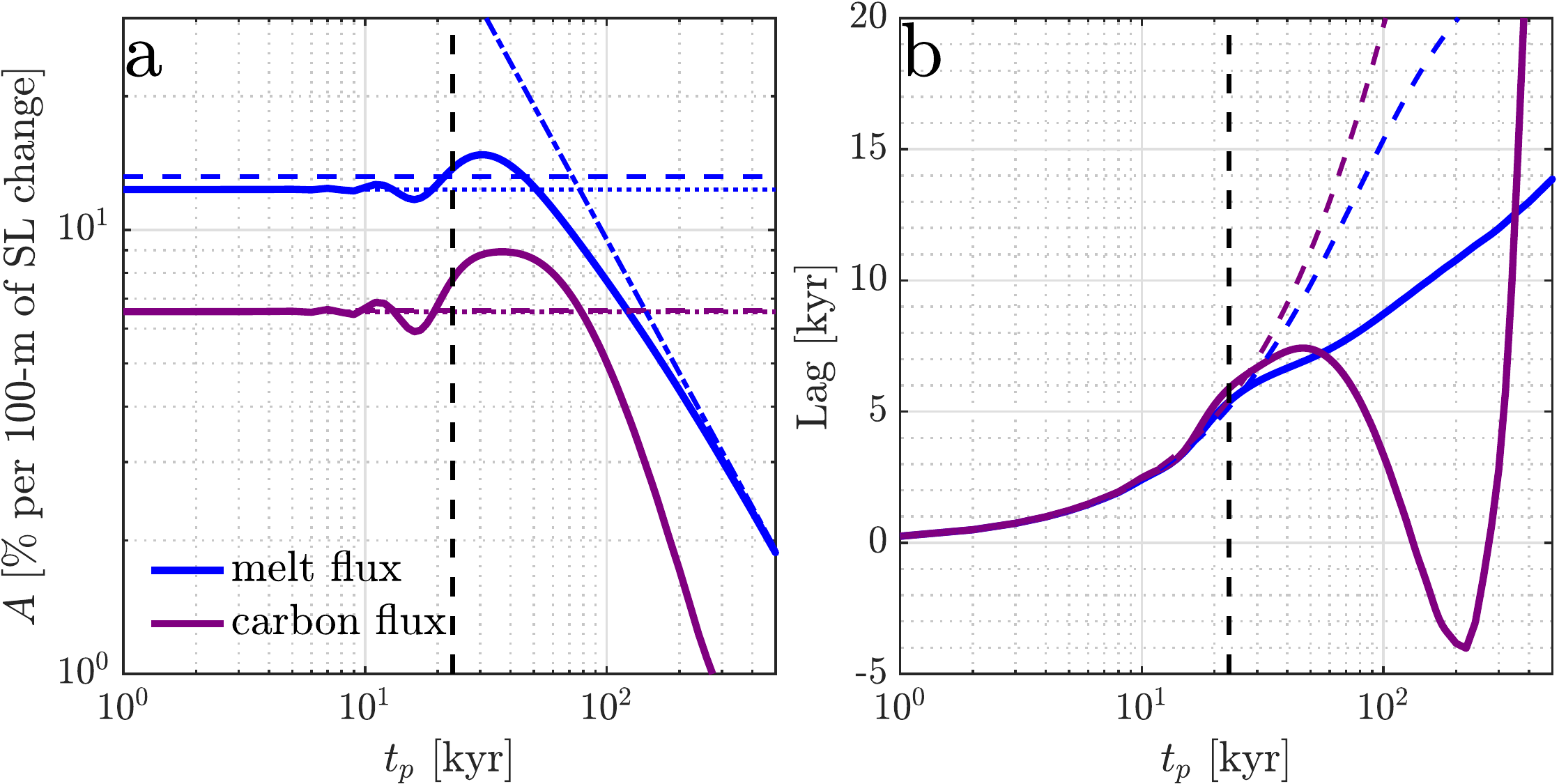}
    \caption{Admittance and lag for models with $\mathcal{Q}=10^{5}$. The definition of parameters and plotted quantities are given in the main text. a) Admittance of melt flux (blue) and carbon flux (purple) as a function of dimensional forcing period $t_p$. The dotted lines are the predictions of admittance at low forcing periods for finite values of $D$ as given by equations~(C.5) and (C.7) for melt and carbon fluxes, respectively. The dashed lines correspond to the approximations assuming $D\ll\overline{\phi}\ll1$ as given by equations~(C.6) and (C.8). The dash-dot line represents the predictions of melt flux admittance at large periods, as given by equation (28). b) Lag of melt flux (blue) and carbon flux (purple) with dimensional forcing period $t_p$. The dashed lines correspond to the approximations given by equations~(C.9a) and (C.9c). In both panels, the vertical black dashed line marks the critical period $t_p^*$; at periods smaller than $t_p^*$ the small-period asymptotic solutions should hold.}
    \label{fig:admittance_wet_log}
\end{figure}

In Appendix C of the main text, we derived a simplified model valid when the forcing period $t_p$ is short compared to a critical period  $t_p^*$.
Figure \ref{fig:admittance_wet_log} demonstrates that the approximate expressions for both the admittance and the lag of fluxes below the critical period $t_p^*$ provide a good estimation of full results of our model.  

In section 3.3 of the main text, we also gave a simple estimate of the admittance of melt flux based on comparing the melting induced by sea-level variation to the background rate of decompression melting. Figure \ref{fig:admittance_wet_log} demonstrates that this estimate is good when the forcing period is much longer than the critical period.

\end{document}